\documentclass[aps,superscriptaddress,prb]{revtex4-2}
\usepackage{amsmath}
\usepackage{graphicx}
\usepackage[T1]{fontenc}

\begin{document}
	\author{Hung-Tzu Chang}
	\altaffiliation{Current address: Max Planck Institute for Biophysical Chemistry, Am Fa\ss berg 11, 37077 G\"ottingen, Germany}
	\affiliation{Department of Chemistry, University of California, Berkeley, CA 94720,
		USA}
	
	\author{Alexander Guggenmos}
	\altaffiliation{Current address: UltraFast Innovations GmbH, Am Coulombwall 1, 85748 Garching, Germany}
	\affiliation{Department of Chemistry, University of California, Berkeley, CA 94720,
		USA}
	
	\author{Christopher T. Chen}
	\affiliation{Molecular Foundry, Lawrence Berkeley National Laboratory, Berkeley,
		CA 94720, USA}
	
	\author{Juwon Oh}
	\altaffiliation{Current address: Department of Chemistry, Soonchunhyang University,
		Asan 31538, Korea}
	\affiliation{Department of Chemistry, University of California, Berkeley, CA 94720,
		USA}
	
	\author{Romain G\'eneaux}
	\altaffiliation{Current address: Universit\'e Paris-Saclay, CEA, CNRS, LIDYL, 91191 Gif-sur-Yvette, France}
	\affiliation{Department of Chemistry, University of California, Berkeley, CA 94720,
		USA}
	
	\author{Yi-De Chuang}
	\affiliation{Advanced Light Source, Lawrence Berkeley National Laboratory, Berkeley,
		CA 94720, USA}
	
	\author{Adam M. Schwartzberg}
	\affiliation{Molecular Foundry, Lawrence Berkeley National Laboratory, Berkeley,
		CA 94720, USA}
	
	\author{Shaul Aloni}
	\affiliation{Molecular Foundry, Lawrence Berkeley National Laboratory, Berkeley,
		CA 94720, USA}
	
	\author{Daniel M. Neumark}
	\email{dneumark@berkeley.edu}
	\affiliation{Department of Chemistry, University of California, Berkeley, CA 94720,
		USA}
	\affiliation{Chemical Sciences Division, Lawrence Berkeley National Laboratory,
		Berkeley, CA 94720, USA}
	
	\author{Stephen R. Leone}
	\email{srl@berkeley.edu}
	\affiliation{Department of Chemistry, University of California, Berkeley, CA 94720,
		USA}
	\affiliation{Chemical Sciences Division, Lawrence Berkeley National Laboratory,
		Berkeley, CA 94720, USA}
	\affiliation{Department of Physics, University of California, Berkeley, CA 94720,
		USA}
	
	\title{Coupled Valence Carrier and Core-Exciton Dynamics
		in WS$_{2}$ Probed by Few-Femtosecond Extreme Ultraviolet Transient Absorption Spectroscopy}
	
\begin{abstract}
	Few-femtosecond extreme ultraviolet (XUV) transient absorption spectroscopy, performed with optical 500-1000 nm supercontinuum and broadband XUV pulses (30-50 eV), simultaneously probes dynamics of photoexcited carriers in WS$_{2}$ at the W O$_3$ edge (37-45 eV) and carrier-induced modifications of core-exciton absorption at the W N$_{6,7}$ edge (32-37 eV). Access to continuous core-to-conduction band absorption features and discrete core-exciton transitions in the same XUV spectral region in a semiconductor provides a novel means to investigate the effect of carrier excitation on core-exciton dynamics. The core-level transient absorption spectra, measured with either pulse arriving first to explore both core-level and valence carrier dynamics, reveal that core-exciton transitions are strongly influenced by the photoexcited carriers. A $1.2\pm0.3$ ps hole-phonon relaxation time and a $3.1\pm0.4$ ps carrier recombination time are extracted from the XUV transient absorption spectra from the core-to-conduction band transitions at the W O$_{3}$ edge. Global fitting of the transient absorption signal at the W N$_{6,7}$ edge yields $\sim 10$ fs coherence lifetimes of core-exciton states and reveals that the photoexcited carriers, which alter the electronic screening and band filling, are the dominant contributor to the spectral modifications of core-excitons and direct field-induced changes play a minor role. This work provides a first look at the modulations of core-exciton states by photoexcited carriers and advances our understanding of carrier dynamics in metal dichalcogenides.
\end{abstract}
\maketitle

\section{Introduction}
Studying the dynamics of elementary excitations in semiconductors such as photoexcited carriers, phonons, and excitons has been crucial to the success
of electronic devices \cite{shahUltrafastSpectroscopySemiconductors1996,Koch2006}. 
While most measurements on the photophysical and photochemical properties of semiconductors are performed within the optical domain, few-femtosecond to attosecond core-level transient absorption (TA) and transient reflectivity spectroscopy have recently been utilized to investigate carrier dynamics in semiconductors and two-dimensional materials \cite{Schultze2014,zurch2017direct,linCarrierSpecificFemtosecondXUV2017,verkampBottleneckFreeHotHole2019,attarSimultaneousObservationCarrierSpecific2020,schlaepferAttosecondOpticalfieldenhancedCarrier2018,buadesAttosecondStateresolvedCarrier2020,britzCarrierspecificDynamics2HMoTe22021} and the decay of core-excitons in insulators \cite{Moulet2017,geneauxAttosecondTimeDomainMeasurement2020,lucchiniUnravellingIntertwinedAtomic2020}. 
Core-level TA spectroscopy in semiconductors typically consists of an optical pulse to excite the carriers in the sample and an extreme ultraviolet (XUV) or X-ray pulse to record the changes in the core-level absorption spectra. In many semiconductors, the core-level absorption spectra can be mapped onto the conduction band (CB) density of states (DOS) due to significant dielectric screening \cite{Rehr2003}, and the core-level TA spectra directly reflects the carrier distributions as a function of energy, thereby providing real-time tracking of carrier dynamics \cite{zurch2017direct,linCarrierSpecificFemtosecondXUV2017,verkampBottleneckFreeHotHole2019,	attarSimultaneousObservationCarrierSpecific2020}. 

In contrast, the core-level absorption spectra of many insulators, in particular ionic solids with poor dielectric screening, exhibit sharp peaks below the onset of core-to-CB edges \cite{excitonNote}. These discrete transitions are termed ``core-excitons'', which are formed by the Coulomb attraction between the excited electron and the core hole \cite{Bassani1980,Hjalmarson1981}. The electron-hole binding results in longer lifetimes of the core-excitons compared to the typical <1 fs decay time of core-to-CB transitions \cite{Strinati1982,grootCoreLevelSpectroscopy2008}. The observation of the decay of core-excitons is enabled through attosecond transient absorption spectroscopy in the extreme ultraviolet, a core-level TA spectroscopy utilizing sub-femtosecond XUV pulses in combination with <5 fs long optical pulses. As opposed to the typical XUV TA measurement in semiconductors where the XUV pulse probes the valence electronic state after optical excitation \cite{Schultze2014,zurch2017direct,zurchUltrafastCarrierThermalization2017,schlaepferAttosecondOpticalfieldenhancedCarrier2018,attarSimultaneousObservationCarrierSpecific2020}, those experiments to probe core-exciton states use the sub-femtosecond XUV pulse to excite the core-excitons, and the core-exciton transition dipoles are subsequently perturbed with the optical pulse \cite{Moulet2017,geneauxAttosecondTimeDomainMeasurement2020,lucchiniUnravellingIntertwinedAtomic2020}, analogous to the studies on the decay of atomic Rydberg states and autoionization states \cite{beckProbingUltrafastDynamics2015}. Due to the large band gap in insulators, which exceeds the photon energy of available visible and ultraviolet light pulses, the observation of the effect of carrier dynamics on a core-excitonic system via core-level TA spectroscopy has not been achieved and the effect of valence electron-hole pairs on core-exciton transitions and their dynamics remains elusive. 

In this work, we report the observation of core-exciton transitions within the W N$_{6,7}$ edge (32-37 eV) of WS$_2$ and the nearby W O$_3$ edge (37-45 eV), a smooth core-level absorption edge consisting of core-to-CB transitions that can be understood within the single-particle mean field picture. The proximity of the two different types of core-level absorption edges presents an excellent opportunity in simultaneously observing the dynamics of carriers in the valence shell and their influence on the dynamics of core-excitons. A single experiment thus probes the carriers at the W O$_3$ edge and the discrete core-exciton transitions at the W N$_{6,7}$ edge.
Tungsten disulfide is a Group VI transition metal dichalcogenide and a semiconducting two-dimensional (2D) layered material. In its mono- and bilayer form, the electronic structure and photophysics of WS$_2$ have been extensively studied for potential applications in optoelectronics, 2D valleytronics and spintronics \cite{Jiang2012a,Berkdemir2013,Gutierrez2013,chernikovExcitonBindingEnergy2014,Ye2014,Sie2014, chernikovPopulationInversionGiant2015,chernikovElectricalTuningExciton2015, rajaCoulombEngineeringBandgap2017,naglerZeemanSplittingInverted2018, rajaEnhancementExcitonPhonon2018,suEnsuremathGammavalleyAssisted2018, guRoomtemperaturePolaritonLightemitting2019}. 
Recently, interlayer charge transfer excitations and novel elementary excitations such as moir\'e excitons were observed in heterostructures containing WS$_2$ layers \cite{Jin2018,Yuan2018,jinObservationMoireExcitons2019,liImagingMoireFlat2021}.

Here, by conducting core-level transient absorption spectroscopy in the XUV on WS$_2$ thin films, picosecond hole relaxation and carrier recombination times are obtained from the core-level TA spectra at the W O$_3$ edge. A $\sim 10$ fs coherence lifetime of core-excitons at the W N$_{6,7}$ edge is also measured. In contrast to the attosecond XUV TA studies on insulators where the observed dynamics are dominated by coupling of core-exciton states with the optical field \cite{Moulet2017,geneauxAttosecondTimeDomainMeasurement2020,lucchiniUnravellingIntertwinedAtomic2020}, here the core-exciton lineshape is primarily influenced by the change of electronic screening and band filling due to carriers excited by the optical pulse. 
The optical-XUV transient absorption study at the W O$_3$ and N$_{6,7}$ edges provides a prototypical example for measuring the carrier-induced modification of core-excitons and paves the way for exploring carrier dynamics in 2D heterostructures and superlattices involving transition metal dichalcogenides, where the element specificity of core-level TA spectroscopy can be employed to enable layer-selective probing of photophysical and photochemical phenomena.

\section{Experiemental Scheme}
The details of sample preparation and the scheme of the XUV TA spectroscopy experiment are provided in Appendix \ref{app:sample} and \ref{app:setup}, respectively. In brief, 40 nm thick 
WS$_2$ films were synthesized on 30 nm thick silicon nitride windows by atomic layer deposition (ALD) of WO$_3$ thin films and subsequent sulfurization with H$_2$S \cite{Kastl2017}. In the XUV TA experiments, the samples were irradiated with a broadband optical pulse (500-1000 nm) with nominal duration of 4 fs and a time-delayed broadband XUV pulse (30-50 eV) produced by high-harmonic generation using a near single cycle optical pulse in a Kr gas jet. 

\section{Results and Discussion}
\begin{figure*}
	\includegraphics[width=.99\textwidth]{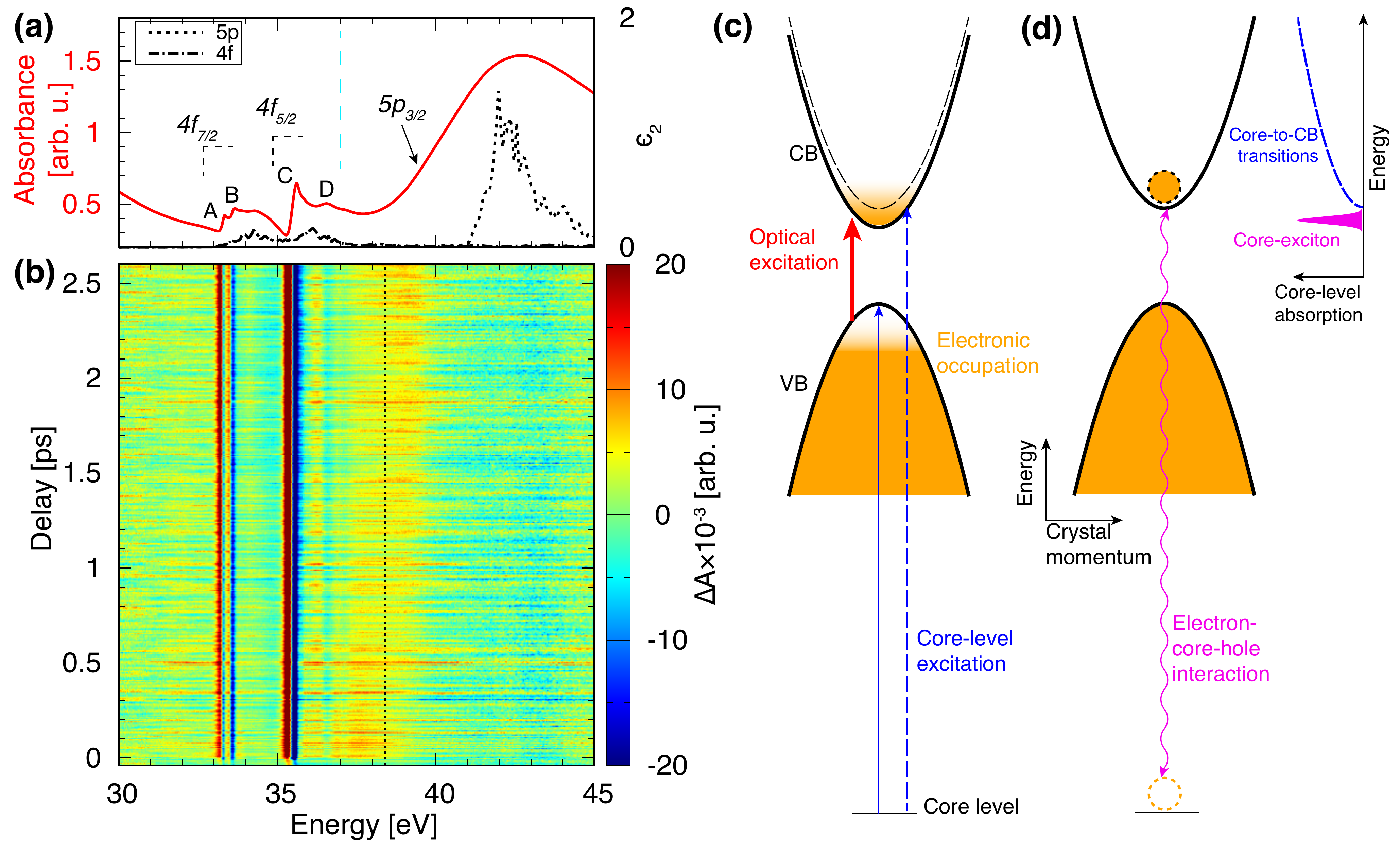}
	\caption{(a) Static core-level absorption spectrum of a 40 nm thick WS$_2$ film (magenta). 
	The dashed and dash-dotted lines show the computed imaginary part of the dielectric function 
	($\epsilon_2(\omega)$) from the $5p$ and $4f$ core bands, respectively. (b) exhibits the XUV TA spectra between -40 fs and 2.6 ps delay. A dashed cyan line and a dotted black line is plotted at 37 eV in (a) and 38.4 eV in (b), respectively, to serve as a reference (see text). The scheme for typical core-level TA measurement in highly screened semiconductors is shown in (c), where the core-level excitation probes the electronic occupation in the VB and CB plus the energy shift of the bands. (d) illustrates the formation of core-exciton through electron-core-hole attraction and its corresponding absorption spectrum.}
	\label{fig:1}
\end{figure*}

The core-level absorption spectrum of the 40 nm thick WS$_2$ film is displayed in Fig.~\ref{fig:1}(a) (red line). 
The static spectrum below 37 eV (marked with dashed cyan line) exhibits four distinct peaks labeled as A-D. Peaks A and B occur on top of the absorption edges between 33 eV and 34 
eV. Peak C exhibits a Fano-type asymmetric lineshape at approximately 35.5 eV with fine structure
peak D occurring at approximately 36.6 eV \cite{Fano1961}. A smooth absorption feature extends from 38.5 eV to beyond 45 
eV. By comparing the measured spectrum with the calculated imaginary part of the dielectric function 
using all-electron full-potential linearized augmented plane wave (FP-LAPW) method (Appendix \ref{app:elk}) with random phase approximation \cite{elk,onidaElectronicExcitationsDensityfunctional2002}, 
the smooth absorption feature above 38.5 eV is assigned to the transition between the W $5p_{3/2}$
core bands and the CB (W O$_3$ edge). Peaks A and B are assigned to the W 
$4f_{7/2}$ transitions (W N$_7$ edge), and peaks C and D to the transitions 
from W $4f_{5/2}$ core-levels to the CB (W N$_6$ edge). The comparison between the measured W 
$5p_{3/2}$ absorption edge and the calculated dielectric function (Fig.~\ref{fig:1}(a), dashed 
line) indicates significant lifetime broadening of the W $5p_{3/2}$-to-CB transitions. In 
addition, the peaks measured at the absorption edges between 32 eV and 37 eV are clearly different from the smooth onset of the $4f$-to-CB transitions (Fig.~\ref{fig:1}(a), 
dash-dotted line) calculated with mean field approximation, suggesting that many-body interactions between the electron and the core hole contribute to the measured discrete lineshape. Note that the broadband XUV pulse (30-50 eV) covers the core-level transitions from both the W $4f$ and $5p$ orbitals, enabling simultaneous observation of dynamics at the two different edges.

Typical core-level TA spectra between -40 fs to 2.6 ps time delay are displayed in Fig.~\ref{fig:1}(b). In this article, the time delay is defined as the arrival time of the XUV pulse subtracted from the time of the optical pulse and positive time delay indicates the samples are probed by the XUV pulse ``after'' optical excitation. 
The change of absorbance $\Delta A(t)$ at a specific time delay $t$ is defined as $\Delta A(t)=A(t)-A(t=-$40 fs$)$ with $A$ denoting absorbance. The pump-excited carrier density is estimated to be $1\times 10^{21}$ cm$^{-3}$ or $6\times 10^{13}$ /(layer$\cdot$cm$^{2}$) (Appendix \ref{app:carrier_density}).
Between 37-40 eV below the W O$_3$ edge, two weak, broad
positive features are observed (Fig.~\ref{fig:1}
(b)). Although the difference of static absorbance below and above the W N$_{6,7}$ edge, or ``edge jump'', is much smaller than the W O$_3$ edge (Fig.~\ref{fig:1}(a)), the XUV TA signal occurring near transitions A, B, and C is narrow and much stronger than the TA signal above 37 eV (Fig. \ref{fig:1}(b)). Clearly, the nature of the W N$_{6,7}$ edge (peaks A-D) and the W O$_3$ edge transitions are different and separate treatment is needed to understand their corresponding XUV TA spectra.

When a core electron is excited into the CB, the excited electron can interact with the core hole via Coulomb attraction. In many semiconductors, the electronic screening reduces the Coulomb interaction such that the core-level transitions can still be understood in a single-particle picture \cite{Rehr2003}. As the core bands are dispersionless, the core-level transitions maps the CB density of states and, with optically excited carriers in the VB and CB, core-level transitions probe the electronic occupation in the valence shell and the energy shifts of the VB and CB due to carrier and phonon excitations (Fig. \ref{fig:1}(c)) \cite{zurch2017direct,
	linCarrierSpecificFemtosecondXUV2017,verkampBottleneckFreeHotHole2019,
	attarSimultaneousObservationCarrierSpecific2020}.
This scheme corresponds to the smooth W O$_3$ edge transitions above 37 eV but cannot describe the transitions at W N$_{6,7}$ edge. 

Discrete peaks form in core-level absorption spectra when the Coulomb attraction between the electron and the core hole is non-negligible. The interaction between the core hole and the electron excited by the XUV or the X-ray renormalizes the core-level absorption and discrete ``core-exciton'' peaks can form near the critical points of the core-to-CB transitions (Fig. \ref{fig:1}(d)) \cite{Bassani1980}. Note that excitonic interactions between the excited electron and the core hole are present for all core-to-CB absorption. Therefore, the renormalization of the core-level absorption spectra is not limited to the near-edge transitions, but may extend several eV above the edge \cite{nakaiNaL2Absorption1969,obrienIntermediateCoupling1991}. The behavior of transitions at the W N$_{6,7}$ edge below 37 eV is consistent with the description of core-excitons. To verify this, we compare the XUV TA spectra of the W N$_{6,7}$ and W O$_3$ edges near zero time delay.

\begin{figure}
	\includegraphics[width=.49\textwidth]{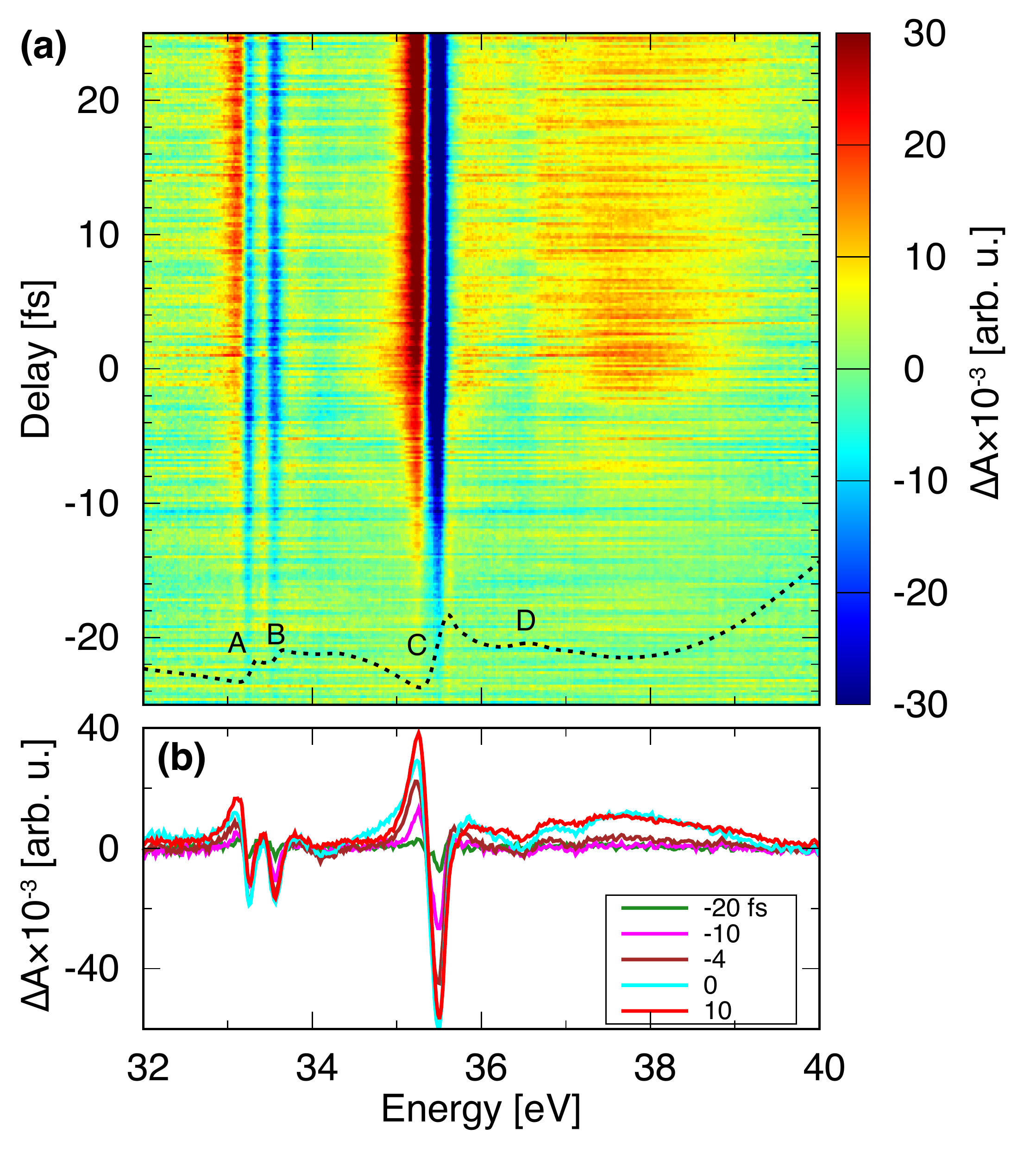}
	\caption{(a) XUV TA signal between -25 fs and +25 fs time delay. The static spectrum is plotted in dotted black line as reference. (b) shows the XUV TA lineouts at 5 different time delays between -20 fs and 10 fs.}
	\label{fig:WS2-short-trace}
\end{figure}

Figure~\ref{fig:WS2-short-trace}(a) displays the XUV TA spectra between -25 fs and +25 fs time delay; lineouts of the XUV TA spectra at 5 different time delays between -20 fs and +7 fs are plotted in Fig. \ref{fig:WS2-short-trace}(b). At positive delays, the XUV light probes the changes due to photoexcitations in the valence shell, while at negative time delays, the optical pulse perturbs the core-level transition dipole before its decay by both carrier photoexcitation and the direct coupling of the core-excitonic transitions with the optical field \cite{Moulet2017,geneauxAttosecondTimeDomainMeasurement2020,lucchiniUnravellingIntertwinedAtomic2020}. The distinction between the transitions at the W N$_{6,7}$ edge and the W O$_3$ edge can be visualized in the XUV TA signal at negative delays. While the TA signal at the W O$_3$ edge (37-40 eV) diminishes to zero at $<-4$ fs time delays, TA signals near peaks A, B, and C are still visible at $\leq -10$ fs time delays. The experimental results thus indicate that the transitions A, B, and C are more long-lived than the transitions below the W O$_3$ edge. The comparison of core-level transition lifetimes corroborates the assignment that the peaks within the W N$_{6,7}$ edge (Fig. \ref{fig:1}(a)) are core-excitons because the electron-core-hole attraction of the core-exciton stabilizes the core-excited state and enables a longer lifetime \cite{Strinati1982}. The broad positive feature above 37 eV at positive time delays can then be interpreted as photoexcited holes in the VB. In the following, we first focus on the measured dynamics induced by photoexcited carriers at the W O$_3$ edge. Next, we discuss the measured TA spectra of core-exciton transitions at the W N$_{6,7}$ edge at both positive and negative time delays. 

\subsection{Carrier dynamics at the W O$_3$ edge}
\begin{figure}
	\includegraphics[width=.49\textwidth]{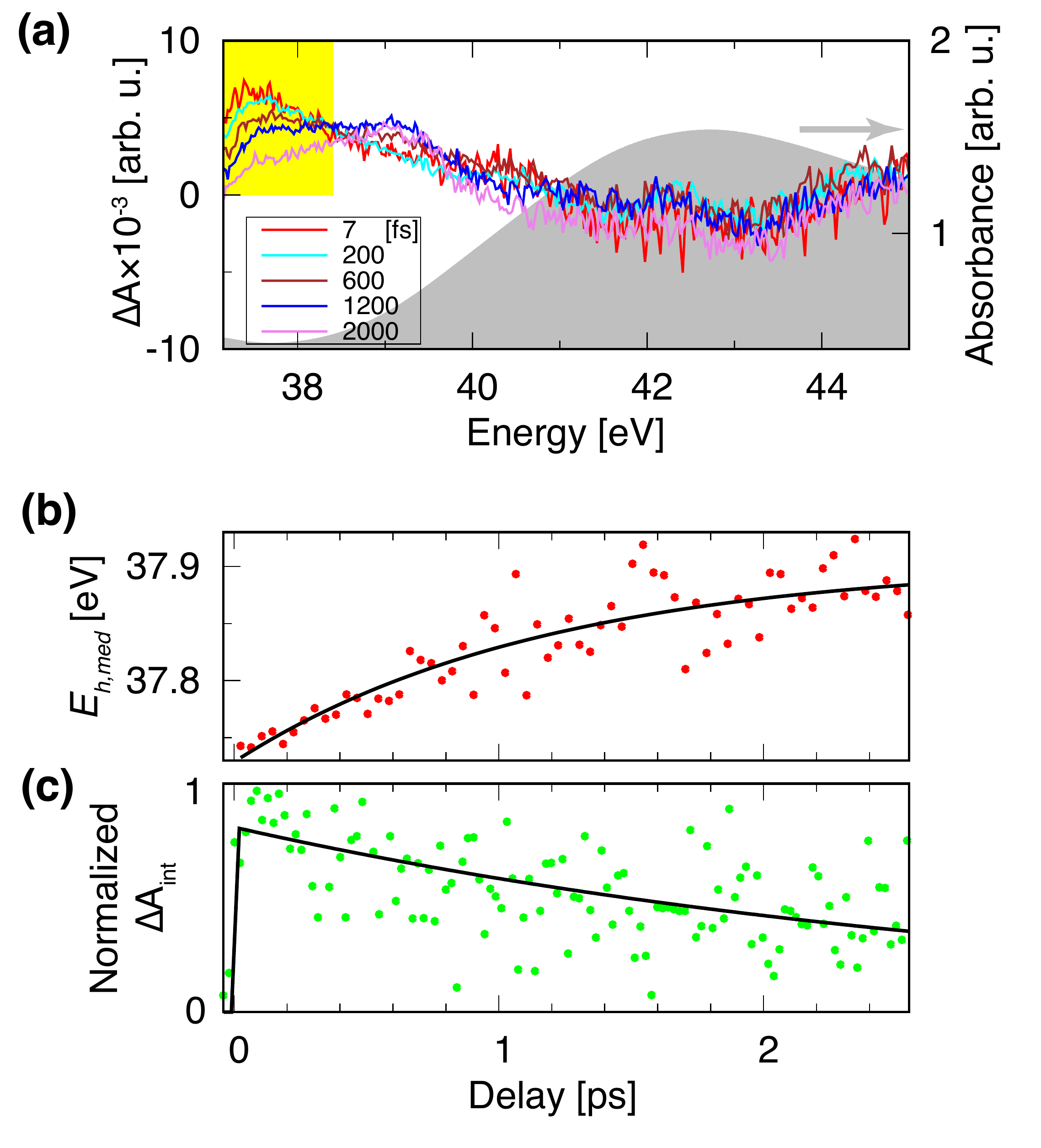}
	\caption{(a) XUV TA spectra at W O$_3$ edge at 5 different time delays. The static core-level absorption spectrum is plotted in gray as reference. (b) shows the median energy $E_{h,med}$ of the TA signal between 37.4--38.4 eV (yellow shaded area in (a)) and (c) shows the integrated XUV TA signal in the same energy region.}
	\label{fig:WS2-O3}
\end{figure}

To understand the XUV TA signal at the W O$_3$ edge, XUV TA spectra at 5 different time delays between +7 fs and +2 ps are presented in Fig. \ref{fig:WS2-O3}(a).  At energies above the edge (41-45 eV), the XUV TA spectra exhibit a weak decrease in absorbance (negative $\Delta A$) throughout the entire range of delays, whereas two positive features with different dynamical behavior are observed between 37-40 eV below the edge. The feature spanning 37-38.4 eV decays with time, while the feature between 38.4 eV and 39.5 eV, which is barely observable near time zero, increases in magnitude with the time delay. 

The positive and negative $\Delta A$ below and above the edge might initially suggest that the positive feature is due to holes in the VB, which open up new excitation pathways from the core, and the negative feature is due to electrons in the CB, which blocks the core-level excitations into the CB. The different dynamical behavior between 37-38.4 eV and 38.4-39.5 eV could then be assigned to relaxation of photoexcited hot hole to the VB edge. However, such an assignment implies that the transition from the core to the VB edge is at approximately 39.5 eV and the initially photoexcited holes are located approximately 2 eV below the VB maximum. Given the approximately 2 eV direct band gap of bulk WS$_2$ \cite{Jiang2012a}, this suggests an optical transition energy of approximately 4 eV, far exceeding the maximum photon energy at 2.5 eV of the optical pulse. Therefore, the positive feature between 38.4-39.5 eV cannot be assigned to holes in the VB. Instead, it can be interpreted as a red shift of the W O$_3$ edge due to band gap renormalization. Both the change in electronic screening and phonon heating due to photoexcited electrons and holes can lead to band gap renormalization, resulting in the lowering of CB edge and therefore a red shift of the core-to-CB transition energies \cite{zurch2017direct}. As energy dissipates from the electronic domain, the increase in phonon temperature and the heat-induced lattice expansion can further enhance band gap renormalization and lowering of the CB \cite{Allen1983}, which causes the positive features to grow with time delay. In addtion, with the assignment of the positive feature spanning 38.4-39.5 eV as due to the phonon-induced edge shift, the feature between 37-38.4 eV, which diminishes with increasing time delay, can then be assigned to holes. 

To track the hole relaxation process, the median energy of the hole signal $E_{h,med}=\frac{\int E\Delta AdE}{\int \Delta AdE}$ is plotted in the inset of Fig. \ref{fig:WS2-O3}(b). The median energy $E_{h,med}(t)$ shifts from 37.7 eV to 37.9 eV with respect to time and can be fitted by a single exponential with a time constant of $1.2\pm 0.3$ ps. In addition, as holes relax to the VB edge at the long time limit, the core-to-VB edge transition energy can be determined from $E_{h,med}(t\to\infty)$. The extracted transition energy from the core to the VB edge from the exponential fitting is 37.9 eV. The proximity between the hole feature and the positive feature due to band gap renormalization leads us to assign the core-to-VB edge transition with the median energy of the hole feature at the long time limit rather than use the maximum energy cutoff of the hole feature as in the core-level TA studies of germanium and 2H-MoTe$_2$ \cite{zurch2017direct,attarSimultaneousObservationCarrierSpecific2020}. 

\begin{figure}
	\includegraphics[width=.49\textwidth]{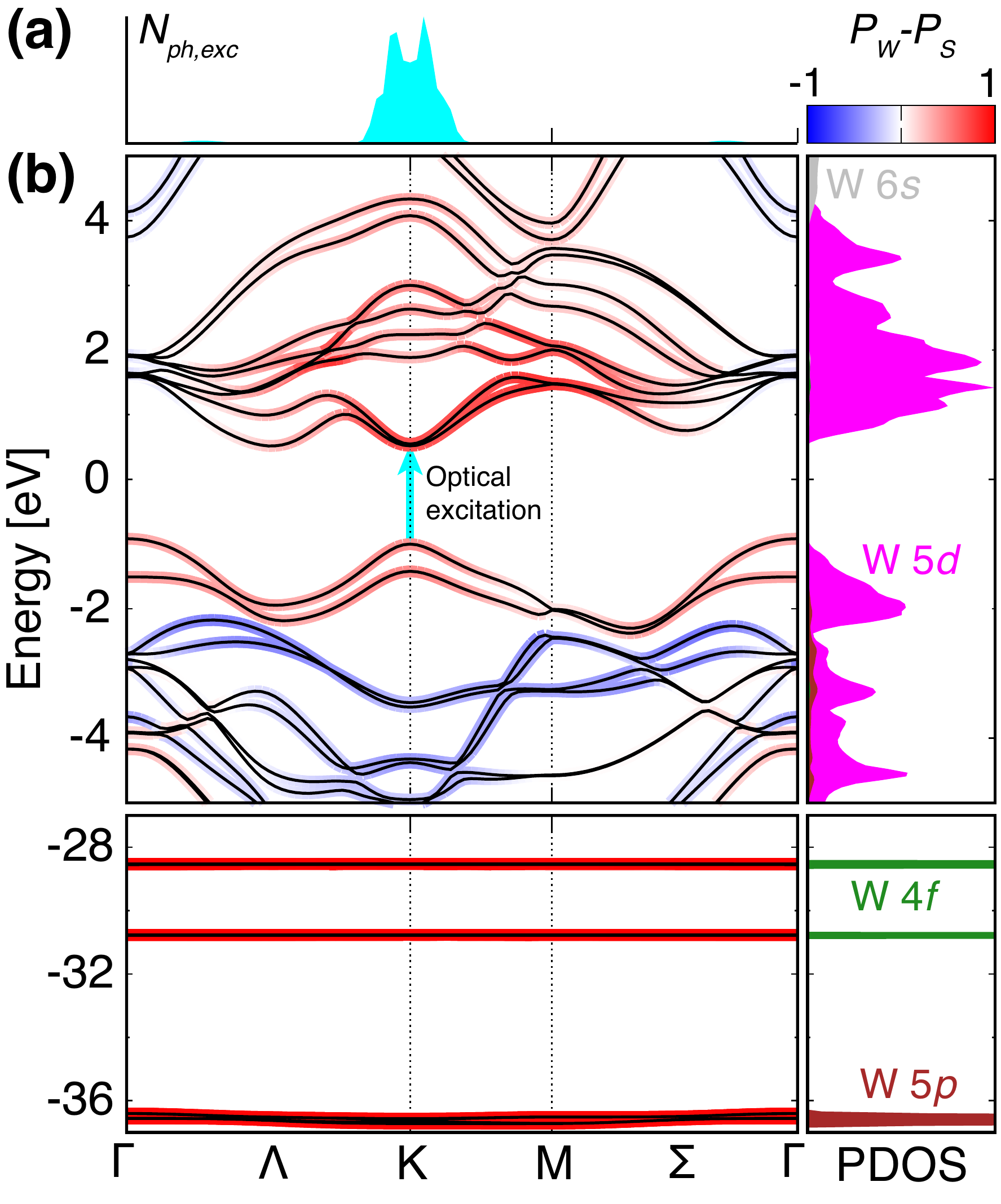}
	\caption{(a) Number of photons in the optical pulse available to excite carriers from the VB to the 
		CB as a function of $k$-points (Appendix \ref{app:elk}). (b) shows the band structure and projected density of states 
		(PDOS) of WS$_2$ calculated with the FP-LAPW method (Appendix \ref{app:elk}). The colors on the bands (left panel) reflects the difference between the 
		atomic orbital projection on W and S ($P_W-P_S$). $P_n$ is the sum of modulus square of 
		wavefunction projections of atom $n$.}
	\label{fig:2}
\end{figure}

Consulting the band structure diagram of WS$_2$ (Fig. \ref{fig:2}(b)) and the initial photoexcited carrier distribution that is proportional to the number of photons available for excitation as a function of crystal momentum (Fig. \ref{fig:2}(a)), the initial carrier distribution is characterized as residing mainly within the K valley and the carriers in the $\Gamma$ valley near the VB maximum are barely excited. The results suggest that the 1.2 ps timescale of hole relaxation is related to the intervalley redistribution of holes between the K valley and the $\Gamma$ valley (VB maximum) mediated by hole-phonon interactions. 

In addition, Fig. \ref{fig:2}(b) shows that the orbital character near the top of VB and bottom of CB is dominated by W $5d$ orbitals, indicating that the core-level transitions from W $4f$ and $5p$ states are both sensitive to carrier dynamics near the band edges. This excludes the possibility that the diminishing TA signal ranging from 37-38.4 eV is due to carriers reaching band regions where core-level transitions from W $4f$ and $5p$ orbitals are forbidden.
The loss of holes in the VB due to recombination is characterized by the integrated TA signal $\Delta A_{int}=\int \Delta A dE$ over the 37-38.4 eV range (Fig. \ref{fig:WS2-O3}(c)). The magnitude of $\Delta A_{int}$ as a function of time can be fitted by a single exponential with a decay constant of $3.1\pm 0.4$ ps convoluted by the instrument response function. The 3.1 ps decay is therefore assigned to the carrier recombination time as the hole signal (37-38.4 eV) decays to zero at the long time limit.

The assignment of the VB maximum at 37.9 eV suggests that a negative TA feature due to excited electrons is expected at 39.9  eV ($37.9+2$ eV band gap). While a weak negative TA signal barely above the noise level is indeed observed spanning 41-45 eV, assigning this to the CB electrons plus the VB maximum at 37.9 eV would suggest an electron-hole energy separation of at least 3 eV, which exceeds the maximum photon energy of 2.5 eV in the optical pulse. In addition, no significant dynamics are observed for the negative feature between 41-45 eV, whereas the electrons are expected to recombine with the holes within the observed 3.1 ps recombination time. This indicates that the weak signal at $\geq$41 eV may be caused by changes in core-hole lifetime, which affects the spectral broadening of the edge, or a decrease in oscillator strength of the W O$_3$ edge due to carrier and phonon excitations \cite{Cushing2018,attarSimultaneousObservationCarrierSpecific2020}, rather than directly due to the photoexcited electrons in the CB.

The absence of the XUV TA signal due to occupation of photoexcited electrons in the CB can be explained by the significant spectral broadening above the W O$_3$ edge, where the experimentally measured static spectrum is much wider and smoother than the calculated dielectric function according to the projected CB DOS (Fig. \ref{fig:1}(a)). Here the fine structure in the projected CB DOS due to critical points in the CB is completely lost in the measured absorption edge. This is in contrast to the recently studied L$_{2,3}$ edge in Si and Te N$_{4,5}$ edge in 2H-MoTe$_2$ where the critical points in the CB can be directly mapped onto the core-level absorption spectrum \cite{Cushing2019,attarSimultaneousObservationCarrierSpecific2020}. The broadening in the W O$_3$ edge increases the overlap between the expected negative XUV TA feature due to electron occupation in the CB and the positive feature due to CB red shift. The overlapping negative and positive features thus lead to the cancellation between the two and make extracting the electron distribution in the CB from core-level TA spectra here unreliable. 

\subsection{Dynamics and lifetimes of core-excitons at the W N$_{6,7}$ edge}
While the core-level TA spectra at the W O$_3$ edge can be explained by the red shift of CB and electronic occupation in the VB and CB, XUV TA signals of core-exciton transitions within the W N$_{6,7}$ edge cannot be interpreted with the same approach. Note that unlike the insulators where core-exciton dynamics have only been observed when XUV light arrives first \cite{Moulet2017,geneauxAttosecondTimeDomainMeasurement2020,lucchiniUnravellingIntertwinedAtomic2020}, here XUV TA signal at the core-exciton transitions extends throughout the entire range of positive time delays (Fig. \ref{fig:1}(b) and Fig. \ref{fig:WS2-short-trace}(a)). This difference arises because in insulators, the band gap exceeds the photon energy range of the optical pulse, so electron-hole pairs in the VB and CB are not excited. Thus, the core-excitons can only be perturbed by optical field induced coupling of core-exciton states. Here, photoexcited carriers with picosecond lifetimes can instantaneously ``dress'' the XUV-excited core-excitons by the carrier-induced change of band filling and electronic screening \cite{leeRoomTemperatureOpticalNonlinearities1986,haugElectronTheoryOptical1984}. Therefore, carrier-induced modification of the core-exciton lineshape occurs at both negative and positive time delays.

In Fig. \ref{fig:WS2-short-trace}(a), the core-level TA signal at short ($<25$ fs) positive delays is solely due to photoexcited carriers since dynamics caused by electron-phonon scattering occur on a timescale of $10^2$ fs and can be ignored \cite{liUltrafastElectronCooling2020,Nie2014,waldeckerMomentumResolvedViewElectronPhonon2017}. 
While photoexcited carriers are the sole contributor to the modification of core-excitons at short positive time delays, at negative delays direct field-induced changes can also modify the core-exciton lineshape. It has been shown that similar to atomic autoionizing states \cite{wangAttosecondTimeResolvedAutoionization2010,Ott2013a,Kaldun2014,beckProbingUltrafastDynamics2015,dingTimeresolvedFourwavemixingSpectroscopy2016}, the optical pulse can cause energy shifts of the core-exciton transitions through the AC Stark effect and resonant coupling between the core-exciton states or with the ionization continuum \cite{Moulet2017,geneauxAttosecondTimeDomainMeasurement2020,lucchiniUnravellingIntertwinedAtomic2020}. Formal treatment of the XUV TA spectra at negative time delays requires computation of the free induction decay of the core-exciton transition dipoles by solving a time-dependent Schr\"odinger equation, including couplings of the core-level transitions with both the photoexcited carriers and the optical field \cite{Ott2013a,Moulet2017,geneauxAttosecondTimeDomainMeasurement2020,lucchiniUnravellingIntertwinedAtomic2020}. However, due to the complexity in including the many-body interactions between the core-level transitions and the photoexcited carriers, which is detailed in the next section, we propose an alternative method to separate the contribution from photoexcited carriers and the optical field by their different time behaviors. 

\begin{figure}
	\includegraphics[width=.48\textwidth]{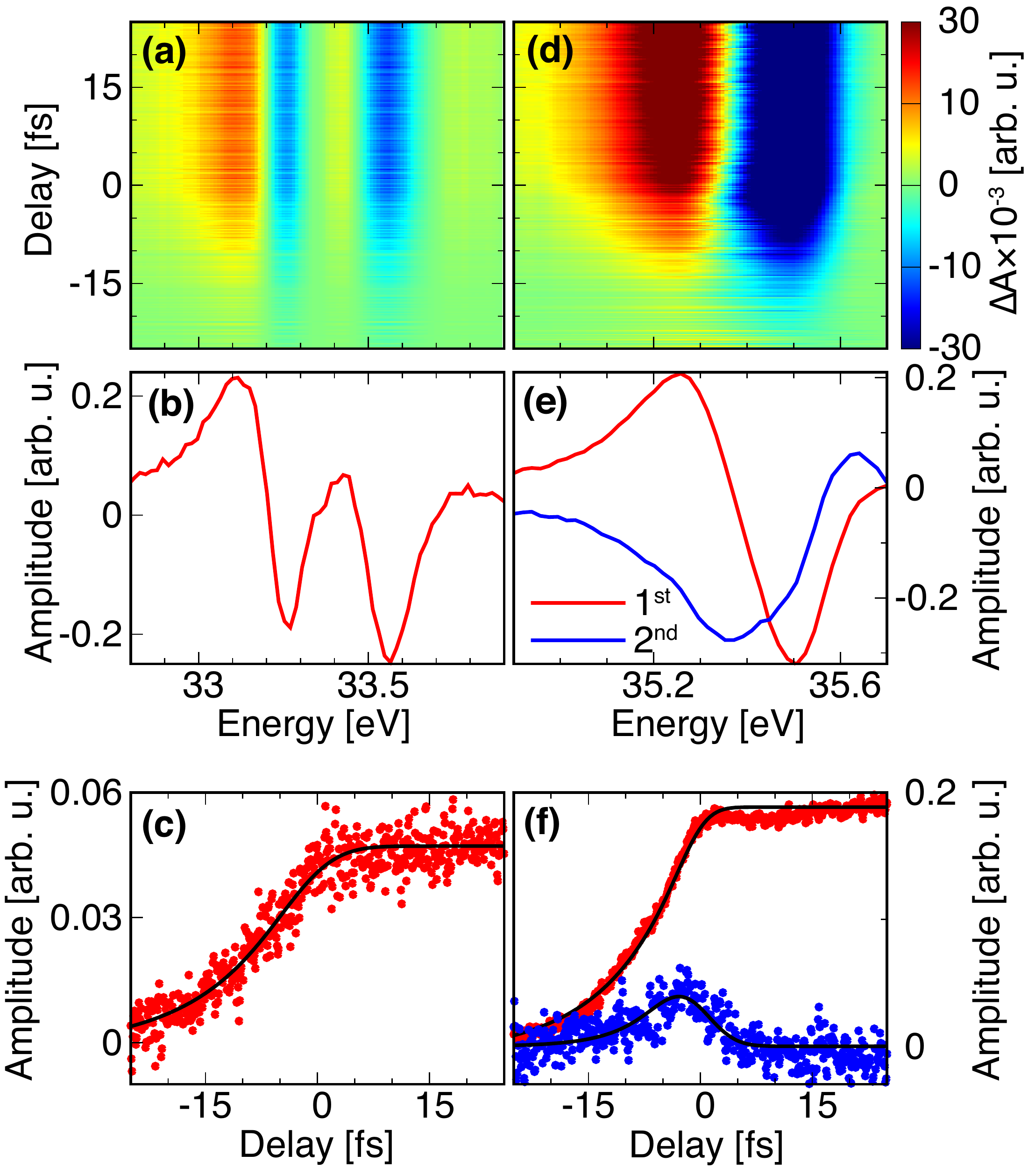}
	\caption{(a) XUV TA signal from the largest component in SVD for transitions A and B and the corresponding singular vector with respect to energy and time are shown in (b) and (c), respectively. (d) exhibits the XUV TA signal from the two highest ranked component in SVD for transition C and the corresponding singular vectors for the first and second largest singular values as a function of energy and time are shown in (e) and (f), respectively. The XUV TA signal from only the largest component in SVD for transition C is shown in Fig. \ref{fig:SVD-C} (Appendix \ref{app:svd}). Note that the magnitude of the singular value of each component is included in the amplitude shown in (c) and (f).}
	\label{fig:SVD}
\end{figure}

Here we analyze the different contributions by applying global fitting to the XUV TA spectra $\Delta A$ through singular value decomposition (SVD) (Appendix \ref{app:svd}): $\Delta A(t,E)=\sum_n s_n u_n(t)v_n(E)$. Functions $\{u_n(t)\}$ and $\{v_n(E)\}$ are singular vectors, or \textit{components}, ranked by singular values $\{s_n\}$ in descending order. Transition D is excluded from the analysis due to poor signal in that spectral region. The XUV TA signal from the largest component at transitions A and B is shown in Fig. \ref{fig:SVD}(a), showing good agreement with experimental data (Fig. \ref{fig:WS2-short-trace}(a)) and indicating the dynamics at transitions A and B can be described by a single component. The corresponding singular vector (Fig. \ref{fig:SVD}(b)) directly reflects the TA signal measured at +10 fs time delay (Fig. \ref{fig:WS2-short-trace}(b)) and the dynamics of the component (Fig. \ref{fig:SVD}(c)) exhibits an exponential decay at negative delay and becomes constant when $t>0$. This indicates that transitions A and B have similar decay dynamics and lifetimes, and although the optical pulse can potentially affect the core-excitons directly through coupling the core-exciton transition dipoles with the optical field, the carriers excited in the valence shell remain the dominant influence on the core-excitons A and B. 

Within their decay time, the core-excitons A and B are modulated by the valence electron-hole pairs. By fitting the decay dynamics with a single exponential (Fig. \ref{fig:SVD}(c)), a core-exciton coherence lifetime ($T_2$) of $10.9\pm0.4$ fs is extracted. The decoherence in core-exciton transitions can be caused by population decay through Auger processes or by exciton-phonon coupling \cite{Moulet2017,geneauxAttosecondTimeDomainMeasurement2020,lucchiniUnravellingIntertwinedAtomic2020}. Previous studies on the decay of core-excitons in insulators show that when exciton-phonon coupling prevails over other decoherence channels, the free induction decay of core-exciton transition dipole moment exhibits a Gaussian decay \cite{Moulet2017,geneauxAttosecondTimeDomainMeasurement2020,lucchiniUnravellingIntertwinedAtomic2020} that leads to a Gaussian spectral profile \cite{huangTheoryLightAbsorption1950,citrinPhononBroadeningXRay1974}. Here the decay of XUV TA signal at negative time delays is exponential rather than Gaussian (Fig. \ref{fig:SVD}(c) and (f); comparison between Gaussian and exponential fitting is detailed in Appendix \ref{app:svd}), suggesting that phonon-induced dephasing is insignificant and Auger processes are the dominant contributor to core-exciton decay. Thus, a population decay time ($T_1\approx T_2/2$) at $5.5\pm 0.2$ fs can be inferred \footnote{As an additional note, while the lifetimes of the core-excitons can be potentially extracted from the core-exciton linewidth in the XUV static absorption spectrum, the overlap between the discrete core-exciton transitions and the continuous core-to-CB transitions (Fig. \ref{fig:1}(a)) makes the linewidth extraction unreliable.}.

To quantitatively reproduce the XUV TA spectra at transition C (Fig. \ref{fig:WS2-short-trace}(a)), the two largest components in the SVD are required (Fig. \ref{fig:SVD}(d)). The dynamics of the two components (Fig. \ref{fig:SVD}(f)) show that at >10 fs delays, the largest (1$^{st}$) component is constant whereas the second component is zero. In addition, the second component only becomes nonzero either at negative time delays or during pulse overlap. This indicates that the largest component represents the influence of carriers on the core-exciton and the second component originates from the direct coupling of the core-exciton to the optical field, because in contrast to the direct coupling to the optical field that can only occur when the field overlaps with the transition dipole before its decay, the carriers are much longer lived than the transient optical pulse and can cause spectral changes at >10 fs time delays.  The field-induced TA component (Fig. \ref{fig:SVD}(e), blue line) exhibits a negative amplitude below the edge and a positive amplitude above (cf. Fig. \ref{fig:3}(b)). The asymmetry of XUV TA amplitude centered around the edge suggests that it may relate to the change in the Fano $q$-factor of the transition, which can be induced by the optical pulse through a ponderomotive phase shift \cite{Ott2013a}, or direct coupling to neighboring core-excited states \cite{Kaldun2014}. Fitting the dynamics of the largest component with a single exponential yields a core-exciton coherence lifetime of $9.6\pm 0.1$ fs. 

\subsection{Carrier-induced modification of core-exciton transitions within W N$_{6,7}$ edge}
In this section, we focus on the modification of core-exciton absorption lineshape by photoexcited carriers. First, we compare the effect of photoexcited carriers on core-exciton and core-to-band absorption at +10 fs time delay (Fig. \ref{fig:WS2-short-trace}(b)). In contrast to the broad, positive TA signal observed below the W O$_3$ edge, no positive signal is observed between 30-33 eV (Fig. \ref{fig:1}(b)) and the majority of the TA signal occurs near the core-exciton transitions A-C. The lack of XUV TA signal directly from the electronic state blocking of carriers as at the W O$_3$ edge can be attributed to the renormalization of core-level absorption spectra due to core-exciton formation. The formation of core-excitons concentrates the oscillator strength of core-to-CB transitions to the bound core-excitonic states and the core-level absorption spectra no longer maps to the CB DOS. In addition, the oscillator strength at the W N$_{6,7}$ edge is lower than the W O$_3$ edge. Compared to the edge change (jump) of 1 OD (optical density) below and above the W O$_3$ edge, the W N$_{6,7}$ edge has an edge jump of $\leq$0.5 OD (Fig. \ref{fig:1}(a)). This indicates that the signal of electronic state blocking, if present at the W N$_{6,7}$ edge, would be much lower than the signal below the W O$_3$ edge (37.4-38.4 eV), which is already close to the noise level. 

\begin{figure*}
	\includegraphics[width=.8\textwidth]{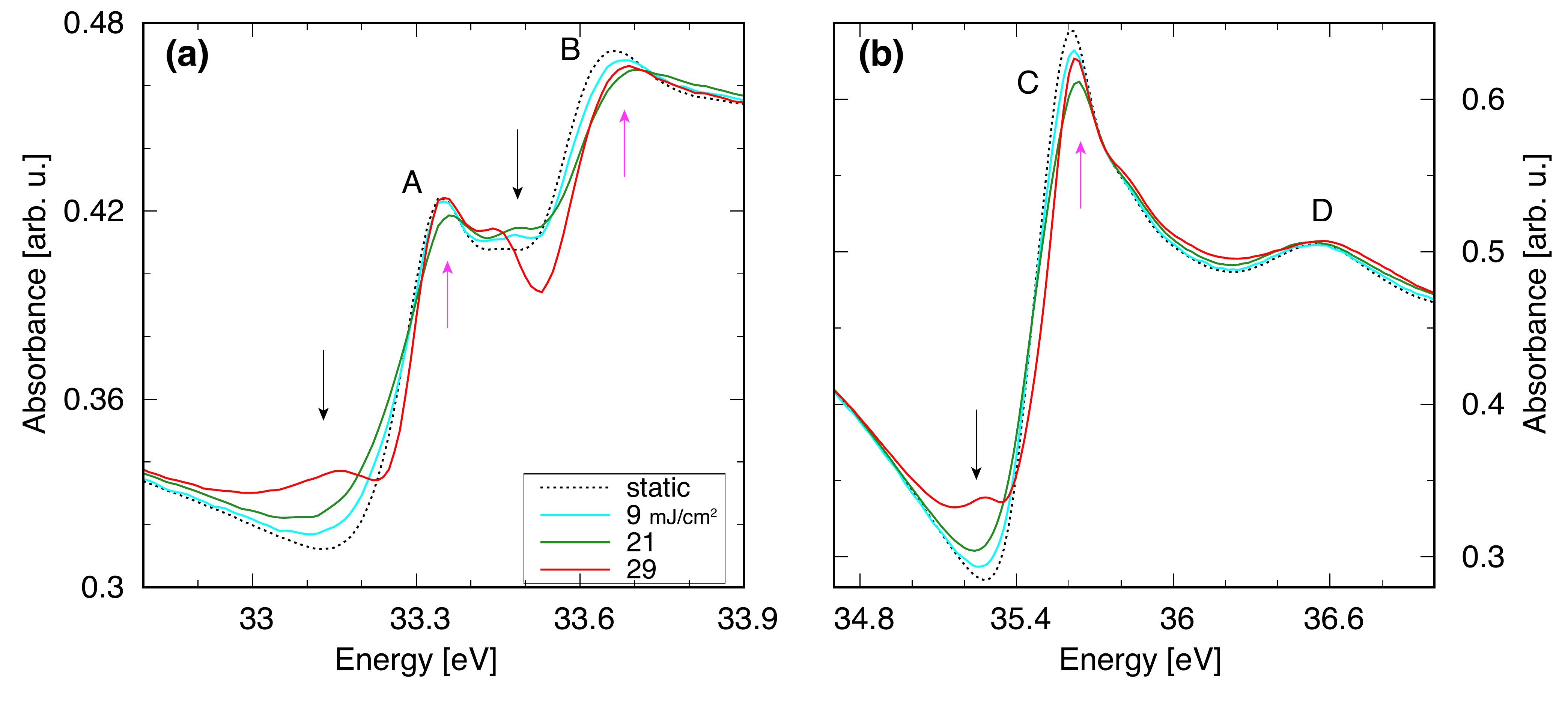}
	\caption{Fluence dependence measurement of core-level absorption spectra at transitions (a) A, B, and (b) C, D, at +10 fs time delay. The static spectrum is shown in black dashed line as a reference. The estimated excited carrier density (Appendix \ref{app:carrier_density}) at pump fluences 9, 21, 29 mJ/cm$^2$ are $4.5\times 10^{20}$, $1.0\times 10^{21}$, and $1.5\times 10^{21}$ cm$^{-3}$, respectively.}
	\label{fig:3}
\end{figure*}

To understand the effect of carriers on core-exciton transitions, the core-level absorption spectra at +10 fs time delay with 3 different optical pump fluences are displayed in Fig.~\ref{fig:3}.
While transition D is broadened with increasing pump fluence, the carrier-induced changes at A, B, and C are much more complex. At transitions A, B, and C, the absorption edge shifts to higher energy with increasing pump fluence. However, the changes in absorbance below and above the edge are non-monotonic with increasing optical excitation. At fluences between 0-21 mJ/cm$^2$, the absorbance below the edges (Fig. \ref{fig:3}, black arrows) increases while the absorbance above the edge (Fig. \ref{fig:3}, magenta arrows) decreases with increasing fluence. At a fluence of 29 mJ/cm$^2$, the absorption above the edge increases rather than decreases compared to the absorbance at 21 mJ/cm$^2$ pump fluence. In addition, shoulders and ripples start appearing around transitions A, B, and C at a fluence of 29 mJ/cm$^2$. New small features appear below the edge of A and C and a dip appears below the absorption edge B. Clearly, the carrier-induced changes to the core-exciton transitions cannot be simply described by an energy shift or a broadening of the lineshape.

The behavior of the core-exciton spectral change at +10 fs delay with respect to increasing pump fluence is reminiscent of the changes in optical absorption of valence excitons in highly excited semiconductors \cite{haugElectronTheoryOptical1984}. For example, it has been observed that in highly excited GaAs, the absorption peak of bound excitons decreases in magnitude and new features below the onset of excitons appear in the optical absorption spectra \cite{leeRoomTemperatureOpticalNonlinearities1986}. Using a generalized Elliot formula \cite{elliottIntensityOpticalAbsorption1957,banyaiSimpleTheoryEffects1986,kochBandEdgeNonlinearities1988}, Lee et al. showed that the contributions to the optical absorption spectra of highly excited semiconductors can be divided into three categories \cite{leeRoomTemperatureOpticalNonlinearities1986}. First, the ionization of bound excitons due to the screening of the photoexcited electron-hole plasma causes suppression of the bound exciton transitions and an overall blue shift of the absorption edge. Secondly, the increasing band filling causes ripples to appear around the exciton transitions, especially below the absorption onset. Thirdly, the carrier-induced band gap renormalization introduces a red shift of the CB edge that partially compensates the blue shift of absorption onset due to suppression of bound exciton transitions. 

The phenomena observed for optical excitons can analogously explain the spectral changes of core-exciton transitions here. The increase of electronic screening due to electron-hole excitations in the VB and CB suppresses bound core-exciton transitions and contributes to the overall blue shift of the absorption onset. The changes of band filling in the valence shell modulates the core-excited states' energies, oscillator strengths, and the coupling to the continuum so that new absorption features appear. 

The similar behavior between the carrier-dressed core-exciton lineshape and optical absorption in highly excited semiconductors, which can be simulated analytically, leads us to explore the possibility of extending the formalism \cite{banyaiSimpleTheoryEffects1986} to quantitatively extract parameters such as core-exciton radii and binding energies. The generalized Elliot formula \cite{elliottIntensityOpticalAbsorption1957,banyaiSimpleTheoryEffects1986,kochBandEdgeNonlinearities1988} is based on a parabolic two-band model that only incorporates a single CB minimum. As core-exciton feature B and D are clearly embedded in the core-to-CB continuum transitions, the parabolic two-band approximation is no longer applicable. In addition, due to the dispersionless core bands, core-excitons can form at multiple CB minima, e.g. at the K, $\Lambda$, and $\Sigma$ valleys (Fig. \ref{fig:2}), and the wavefunctions at those CB minima can further hybridize. Therefore, quantitative treatment of core-exciton transitions here, and by extension, their modification due to carriers, will require Bethe-Salpeter equation calculations including the full bandstructure of WS$_2$ \cite{haugDynamicalScreeningExcitons1978,zimmermannDynamicalScreeningSelfenergy1978,schmitt-rinkManybodyEffectsAbsorption1986}, which is beyond the scope of this work.

\subsection{Picosecond XUV transient absorption signal at the W N$_{6,7}$ edge}
\begin{figure}
	\includegraphics[width=.49\textwidth]{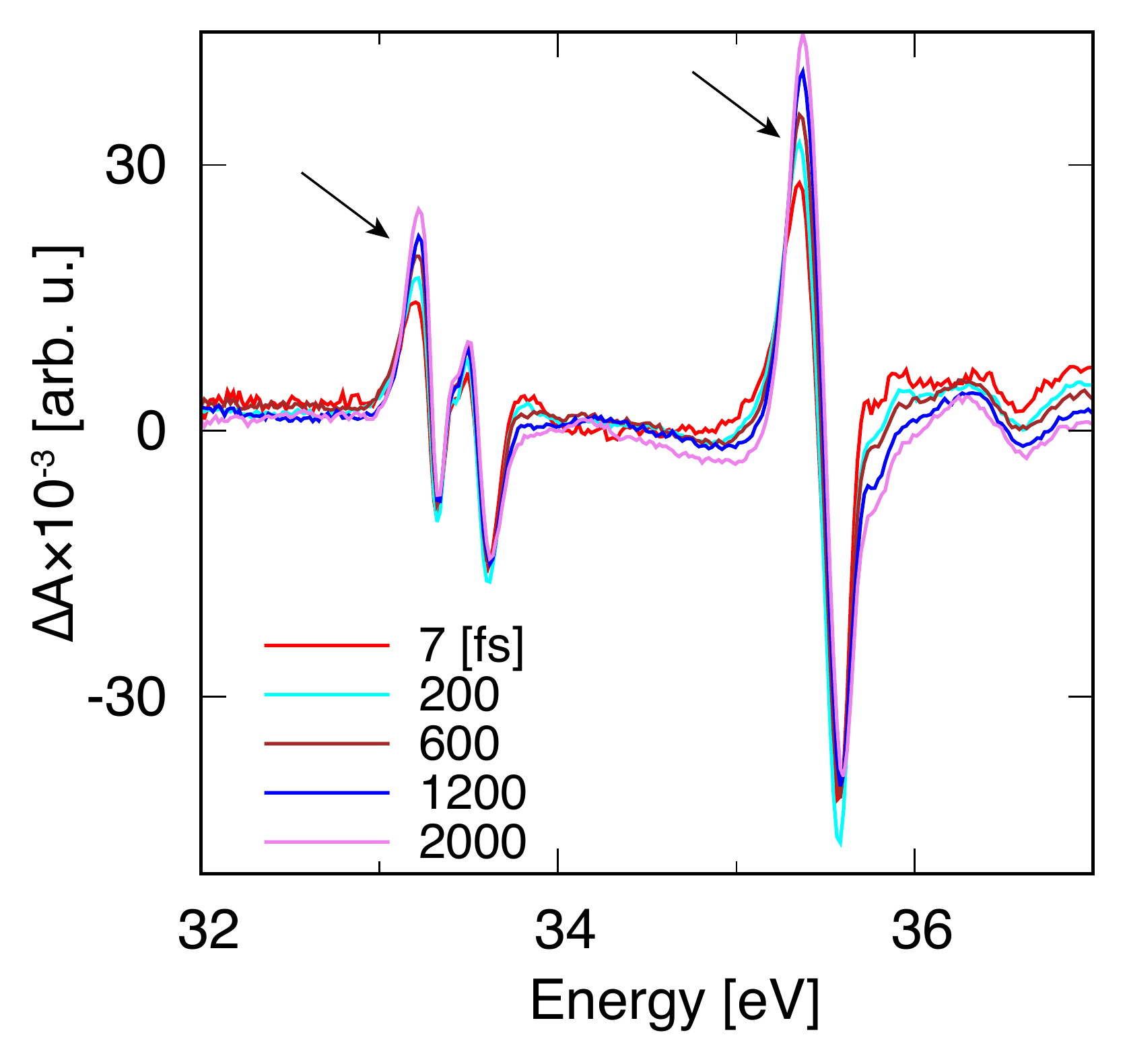}
	\caption{XUV TA spectra at 5 different positive time delays between 0-2 ps.}
	\label{fig:WS2-core-exciton-long}
\end{figure}

As photoexcited carriers are the major contributor to the modulations of the core-exciton spectra at negative and short positive time delays, we consider here the possibility of using the TA spectra of core-excitons to extract carrier dynamics. Although picosecond carrier relaxation and recombination would suggest a decay of TA signal at the core-exciton transitions, a growth of TA signal (Fig. \ref{fig:WS2-core-exciton-long}, black arrows) is observed below the transitions A and C with increasing time delay and no significant TA change is measured above the edge at transitions A, B, and C throughout 0-2.6 ps. This indicates that in addition to photoexcited carriers, the excitation of phonons through electron-phonon interactions also contribute to the spectral changes of core-excitons at long time delays, as phonon induced band gap renormalization can induce a red shift of CB that is consistent with the positive TA signal observed below the transitions. Therefore, the core-exciton transitions at W N$_{6,7}$ edge here are poorly configured for extraction of carrier dynamics, because the spectral changes due to carriers and phonons at hundreds-of-femtoseconds to picosecond timescales cannot be easily separated.

\subsection{Comparison between the core-exciton and core-to-conduction band transitions}
The contrasting behavior of core-exciton transitions at the W N$_{6,7}$ edge and the core-to-band transitions at the W O$_3$ edge in the same energy range is highly unique and suggests that factors other than the macroscopic screening, which are experienced by transitions from both W $4f$ and $5p$ core-levels, are contributing to the core-exciton formation. As the CB minima are dominated by W $5d$ orbitals that are accessible from both W $4f$ and $5p$ through XUV photons, it is thus suggested that the contributor to the difference between the W N$_{6,7}$ edge and the W O$_3$ edge absorption lies in the properties of the core orbitals. The W $4f$ orbitals involved at the W N$_{6,7}$ edge are far more localized than the W $5p$ orbitals for the O$_3$ edge transitions. The localized core hole may then act as a point positive charge and modulate the electronic wavefunctions in the CB to form a core-exciton \cite{Hjalmarson1981}. 

\section{Conclusion}
In summary, photoinduced dynamics at W N$_{6,7}$ and O$_3$ edges in WS$_2$ are simultaneously measured by XUV core-level transient absorption spectroscopy. Picosecond hole relaxation and recombination dynamics in the valence band are extracted from the transient absorption spectra of the core-to-conduction band transitions at the W O$_3$ edge. Lifetimes of core-excited states at the W O$_3$ edge and the W N$_{6,7}$ edge are obtained from XUV transient absorption spectra at negative time delays. While the lifetimes of W O$_3$ edge transitions are well below the duration of the optical pulse ($\sim 4$ fs), core-exciton coherence lifetimes up to 11 fs are observed at the W N$_{6,7}$ edge. Global fitting of the XUV transient absorption spectra at short time delays reveals that in contrast to the direct field-induced core-exciton dynamics observed in insulators \cite{Moulet2017,geneauxAttosecondTimeDomainMeasurement2020,lucchiniUnravellingIntertwinedAtomic2020}, carrier-induced modulation of core-exciton states dominates the dynamics at the few-tens-of-femtosecond timescale. 

The drastically different behavior between the absorption from the W $5p$ and $4f$ core orbitals in the same energy region suggests that in addition to macroscopic screening, the degree of localization of the core orbitals can contribute significantly to the core-level absorption lineshape and the formation of core-excitons. The observation of carrier-modulated core-exciton transitions can serve as an initial step in further understanding and manipulating the dynamics of core-excitons in condensed matter, and the extraction of hole dynamics at W O$_3$ edge further advances the use of core-level TA spectroscopy in measuring carrier dynamics in transition metal dichalcogenides and their heterostructures.

\appendix
\section{Sample preparation} \label{app:sample}
The WS$_2$ sample was synthesized by atomic layer deposition of WO$_3$ thin films on 30 nm thick silicon nitride membranes (Norcada Inc.). The tungsten oxide film was subsequently converted sulfide in a tube furnace with H$_2$S. Before atomic layer deposition, 16 nm thick silicon nitride films were deposited onto the Si frame of the silicon nitride windows using plasma-enhanced chemical vapor deposition (PECVD) to prevent silicon sulfide formation during the reaction with H$_2$S \cite{doeringHandbookSemiconductorManufacturing2008}. The passivated windows were then coated with WO$_3$ using atomic layer deposition in an oxygen plasma \cite{Kastl2017}. The thickness of WO$_3$ was calculated from the required thickness of WS$_2$ using the ratio of the density between the two assuming no W loss in the reaction with H$_2$S. The thickness of the oxide film was characterized by \textit{in situ} spectroscopic ellipsometry. After the oxide deposition, the windows were put in a quartz boat and transferred into a tube furnace which was heated up to 600 \textdegree C. H$_2$S (5 sccm) and Ar (100 sccm) as a buffer gas was flowed into the tube to react with WO$_3$. After 1 hr of reaction, the H$_2$S flow was turned off while maintaining the Ar flow to prevent contamination from outside air and the furnace is left to cool down. After the temperature reached below 200 \textdegree C, the Ar flow was switched to N$_2$ and the samples were taken out after the instrument reached room temperature. To verify that the absorption peaks below 37 eV are not due to defect-induced color centers, the XUV absorption spectrum of the synthesized film is compared with total electron yield (TEY) spectrum of single crystal WS$_2$ (2Dsemiconductors USA) measured at Beamline 4.0.3 at the Advanced Light Source (Appendix \ref{app:ALS}). 

\section{Experimental setup} \label{app:setup}
\begin{figure}
	\includegraphics[width=.49\textwidth]{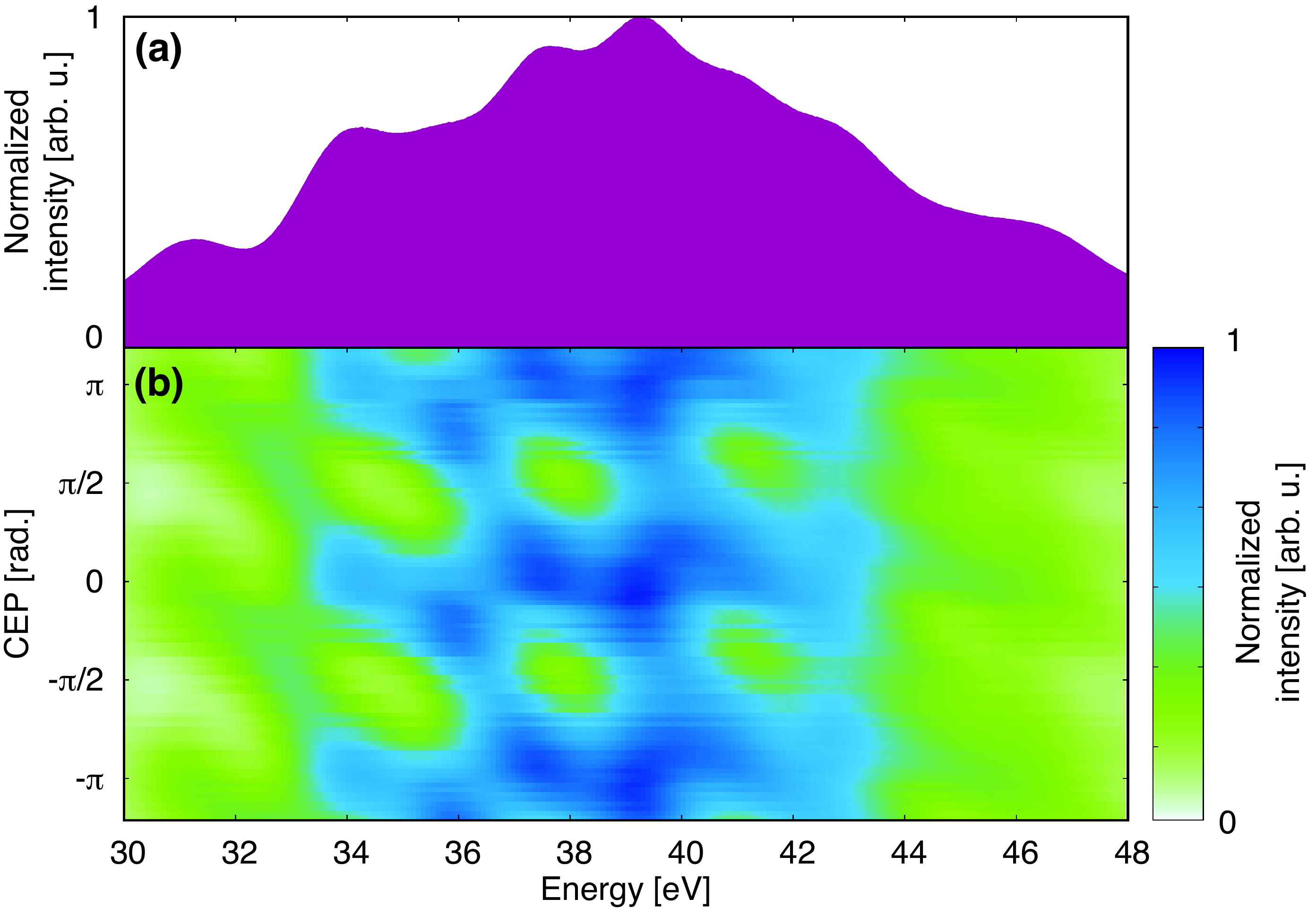}
	\caption{(a) A typical XUV spectrum produced by high-harmonic generation in Kr and (b) the XUV spectra as a function of carrier envelope phase of the driving pulse}
	\label{fig:XUV_pulse}
\end{figure}

\begin{figure*}
	\includegraphics[width=.98\textwidth]{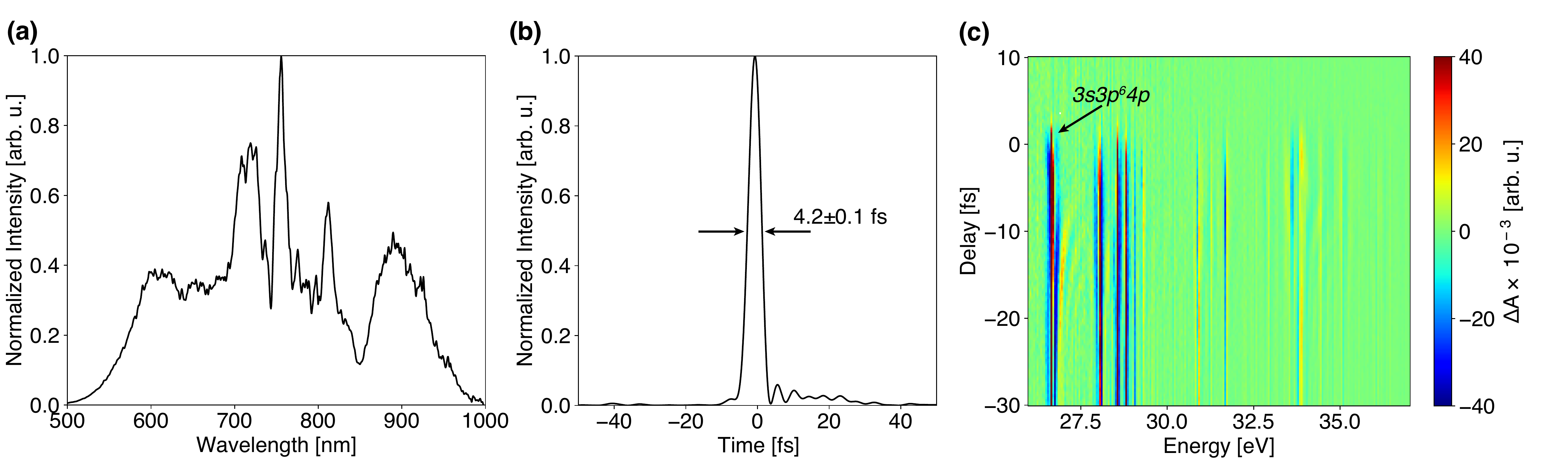}
	\caption{(a) Intensity calibrated spectrum and (b) temporal profile of the pump pulse. (c) transient absorption spectra of Ar $3s3p^6np$ autoionization states for time zero calibration.}
	\label{fig:pulse_characterization}
\end{figure*}

The optical and XUV pulses in the experiment was produced by a Ti:sapphire carrier envelope phase (CEP) stabilized laser operating at 1 kHz (Femtopower Compact Pro seeded by Femtolaser Rainbow CEP3). The output of the Ti:sapphire laser was 1.8 mJ in pulse energy and approximately 30 fs in pulse duration. The laser beam was focused into a 1 m long Ne-filled hollow core fiber to generate a supercontinuum spanning 500-1000 nm wavelength with self-phase modulation. A mechanical chopper was installed after the hollow core fiber to chop down the repetition rate to 100 Hz to prevent sample damage through excessive heating. The dispersion accumulated during pulse propagation was compensated by a set of broadband double-angle chirped mirrors (PC70, UltraFast Innovations) and a 2 mm thick ammonium diphosphate crystal \cite{Timmers2017}. The beam was then separated into the probe and pump arm by a 9:1 broadband beamsplitter. Each arm was equipped with a pair of UV-graded fused silica wedges for dispersion fine-tuning. The probe beam was subsequently focused into a Kr gas jet to produce broadband XUV pulses (30-50 eV) via high-harmonic generation (Fig. \ref{fig:XUV_pulse}). The XUV beam then traveled through a 100 nm thick Al filter blocking the high-harmonic driving field and is focused onto the sample with a Au coated toroidal mirror. The pump beam was time-delayed with respect to the probe by a piezo-driven optical delay stage and was subsequently recombined with the probe arm by an annular mirror. A 200 nm thick Al filter is placed after the sample to prevent the pump beam reaching the XUV spectrometer. The XUV beam passing through the sample and the Al filter was dispersed by a flat-field grating onto an XUV CCD camera. The spectral energies were calibrated with the autoionization lines of Ar $3s3p^6np$ and Ne $2s2p^6np$ states \cite{codlingResonancesPhotoionizationContinuum1967,maddenResonancesPhotoionizationContinuum1969}. The duration of the pump pulse was characterized by dispersion scan \cite{Silva2014} to be $\tau_{pump}=4.2\pm 0.1$ fs and the spectrum and temporal profile of the pump pulse are shown in Fig. \ref{fig:pulse_characterization}(a) and (b), respectively. The pulse energy of the pump beam was controlled by an iris and the beam profile of the pump pulses was imaged directly at the sample position with a CMOS camera to calculate the pump fluence. During the XUV transient absorption experiment, the sample was raster-scanned to prevent heat damage.

To avoid the drift of time delay during the experiments, an optical-XUV transient absorption measurement on Ar was conducted after each WS$_2$ transient absorption scan \cite{zurch2017direct,attarSimultaneousObservationCarrierSpecific2020}. The Ar gas cell was mounted alongside the WS$_2$ sample. The suppression of Ar $3s3p^6np$ autoionization lines by the optical pulse at 26-37 eV photon energies was measured (Fig. \ref{fig:pulse_characterization}(c)) \cite{maddenResonancesPhotoionizationContinuum1969,wangAttosecondTimeResolvedAutoionization2010} and the time reference of each scan was determined by fitting the integrated absolute value of transient absorption signal of the Ar $3s3p^64p$ state along the energy axis and fit it with a Gaussian error function \cite{zurch2017direct}. The time axis of each WS$_2$ transient absorption scan was shifted according to its time zero reference and the transient absorption signal $\Delta A$ interpolated onto a uniform time delay grid. In addition to time zero referencing, the cross-correlation time between the pump and probe pulses was estimated by the width of the error function rise \cite{zurch2017direct,wangAttosecondTimeResolvedAutoionization2010} and the estimated pump-probe cross-correlation time is $\tau_{cc}=4.1\pm 0.5$ fs. The maximal cross-correlation time of the 4.2 fs pump pulse and a single-cycle driving pulse for high-harmonic generation centered at 730 nm (Fig. \ref{fig:pulse_characterization}(a)) is $\sqrt{4.2^2+2.4^2}\approx 4.8$ fs. The experimentally measured cross-correlation time of $4.1\pm 0.5$ fs is within the cross-correlation of the pump pulse and a single-cycle driving pulse, and the maximum width of the XUV pulse train envelope is estimated to be $\sqrt{(\text{max}\:\tau_{cc})^2-(\text{min}\:\tau_{pump})^2}\approx 2.1$ fs, less than two half cycles of the optical driving field. This indicates that the XUV pulse train consists of $\leq 2$ attosecond XUV bursts.

To provide a reference for future studies on carrier effects on core-excitons in solids that cannot be prepared as thin films, we performed XUV transient reflectivity experiments on 40 nm thick WS$_2$ thin films deposited on silicon wafers, which were synthesized alongside the samples for XUV transient absorption experiments (Appendix \ref{app:sample}). The measurements were taken on an almost identical beamline as the one for XUV transient absorption, except for the interaction geometry at the sample \cite{Kaplan2018}. The optical pump and XUV probe pulses (p- and s-polarized, respectively) impinged on the sample surface with a $66^{\circ}$ angle from the sample normal. The reflected XUV beam was directed into an identical spectrometer as the one used in absorption. A gold mirror was used as a reference to extract absolute reflectivity of the WS$_2$ sample \cite{Kaplan2018} and because of the relatively weak change in reflectivity, the data was processed using edge-pixel referencing \cite{Geneaux2021}. The results of the XUV reflectivity measurements are detailed in Appendix \ref{app:reflec}.

\section{Electronic structure calculations} \label{app:elk}
The electronic structure of bulk WS$_2$ is computed with all-electron full-potential linearized augmented plane wave (FP-LAPW) method using the Elk code \cite{elk,singhPlanewavesPseudopotentialsLAPW2006}. The density functional theory (DFT) computation is conducted within local spin density approximation (LSDA) \cite{perdewAccurateSimpleAnalytic1992}. Spin-orbit coupling effects are included and the calculations are converged with a $k$-grid of $10\times 10\times 3$ $k$-points. A 4 eV blue shift is added to the calculated dielectric function in Fig. \ref{fig:1}(a) to compensate the underestimated gap between the core-levels and the CB in DFT calculations.

The number of photons available in the optical pulse to excite valence electron-hole pairs as a function of $k$-points (Fig. \ref{fig:2}) is calculated with the formula 
$$ N_{ph,exc}(\mathbf{k})=\sum_{n_v,n_c}\int dE\: N_{ph}(E)\delta (E_{n_c}(\mathbf{k})-E_{n_v}(\mathbf{k})).$$
The number of photons as a function of photon energy $N_{ph}(E)$ is obtained from the measured spectrum of the pump pulse (Fig. \ref{fig:pulse_characterization}(a)). The energies of valence and conduction bands $n_v$ and $n_c$ are obtained from the calculated band structure (Fig. \ref{fig:2}(b)).

\section{Optically excited carrier density} \label{app:carrier_density}
The photoexcited carrier density $\rho_{exc}$ is estimated by calculating the number of absorbed photons in the 40 nm thick WS$_2$ film per unit area $\sigma_{abs}$ divided by the thickness of the film $d$: $\rho_{exc}=\sigma_{abs}/d$. The number of photons absorbed per unit area can be calculated with the equation
$$\sigma_{abs}=\int d\omega\: \bar{\sigma}_{inc}(\omega)f_{abs}(\omega),$$
where $\bar{\sigma}_{inc}(\omega)$ is the number of incident photons per unit area with photon energy $\omega$ and $f_{abs}$ is the fraction of photons absorbed in the film. $\bar{\sigma}_{inc}(\omega)$ can be calculated from the spectrum of the pump pulse and the measured fluence. The fraction of photons absorbed ($f_{abs}$) is calculated using the transfer matrix method including the 40 nm thick WS$_2$ film and the silicon nitride window \cite{burkhardAccountingInterferenceScattering2010}. The refractive indices of WS$_2$ and silicon nitride are taken from Ref. \cite{hsuThicknessDependentRefractiveIndex2019} and \cite{vogtDevelopmentPhysicalModels2015}. 
We conducted the experiments with fluences ranging 6-30 mJ/cm$^2$ and the resulting calculated excited carrier density ranges $3\times 10^{20}$-$2\times 10^{21}$ cm$^{-3}$. The carrier density per layer is calculated multiplying the carrier density by volume with the layer thickness of 6.2 \AA~\cite{Villars2016:sm_isp_sd_0551013}.

\section{Singular value decomposition} \label{app:svd}
The XUV TA spectra between -25 fs and 25 fs below 37.5 eV are analyzed with global fitting via singular value decomposition (SVD), where the TA signal $\Delta A(t,E)$ is written as a matrix with rows and columns indicating different time $t$ and energy $E$, respectively. The TA matrix $\Delta A(t,E)$ is then decomposed with SVD into $\Delta A(t,E)=U(t)^TSV(E)$, where $U$ and $V$ are unitary matrices consisting of singular vectors $\{u_n(t)\}$ and $\{v_n(E)\}$, respectively. $S$ is a rectangular diagonal matrix and the diagonal matrix elements $S_{nn}=s_n$ are singular values ranked in descending order. The reconstruction of TA signal $\Delta A_{rec}(t,E)$ by components up to the $n^{th}$ rank is defined as $\Delta A_{rec}(t,E)=\sum_{m=1}^n s_m u_m(t) v_m(E)$. Note that the SVD approach is based on the assumption that the transient absorption spectra can be represented by a linear combination of components from different contributions. Here such an assumption is valid because the carrier-induced spectral modification is the dominant contributor (Fig. \ref{fig:SVD}(a)-(c)) and the direct field-induced effects can be regarded as a minor component. 
\begin{figure*}
	\includegraphics[width=.98\textwidth]{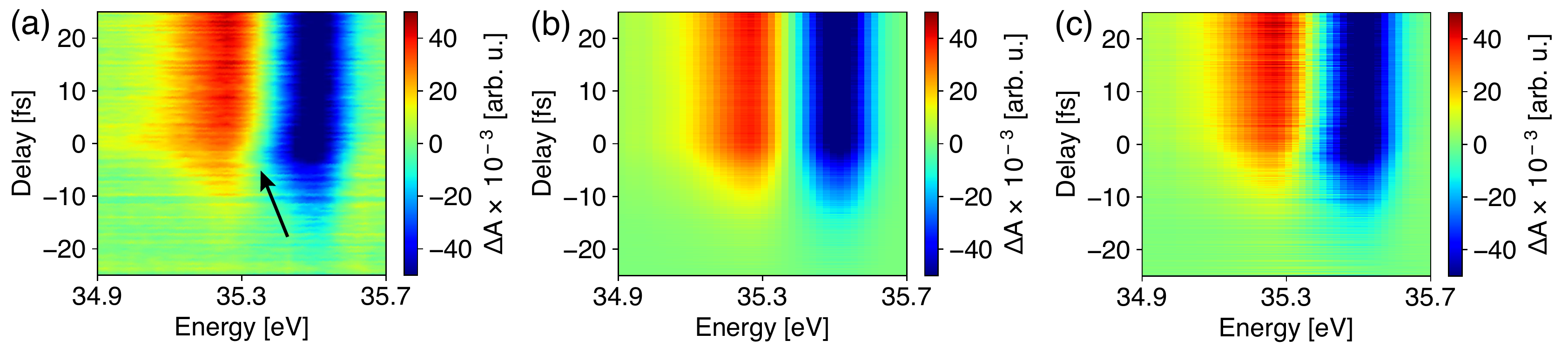}
	\caption{(a) Experimental short time XUV TA spectra at transition C. (b) shows the reconstruction of the XUV spectra with the first (largest) spectral component and (c) shows the reconstruction with the first and second component.}
	\label{fig:SVD-C}
\end{figure*}

\begin{figure}
	\includegraphics[width=.49\textwidth]{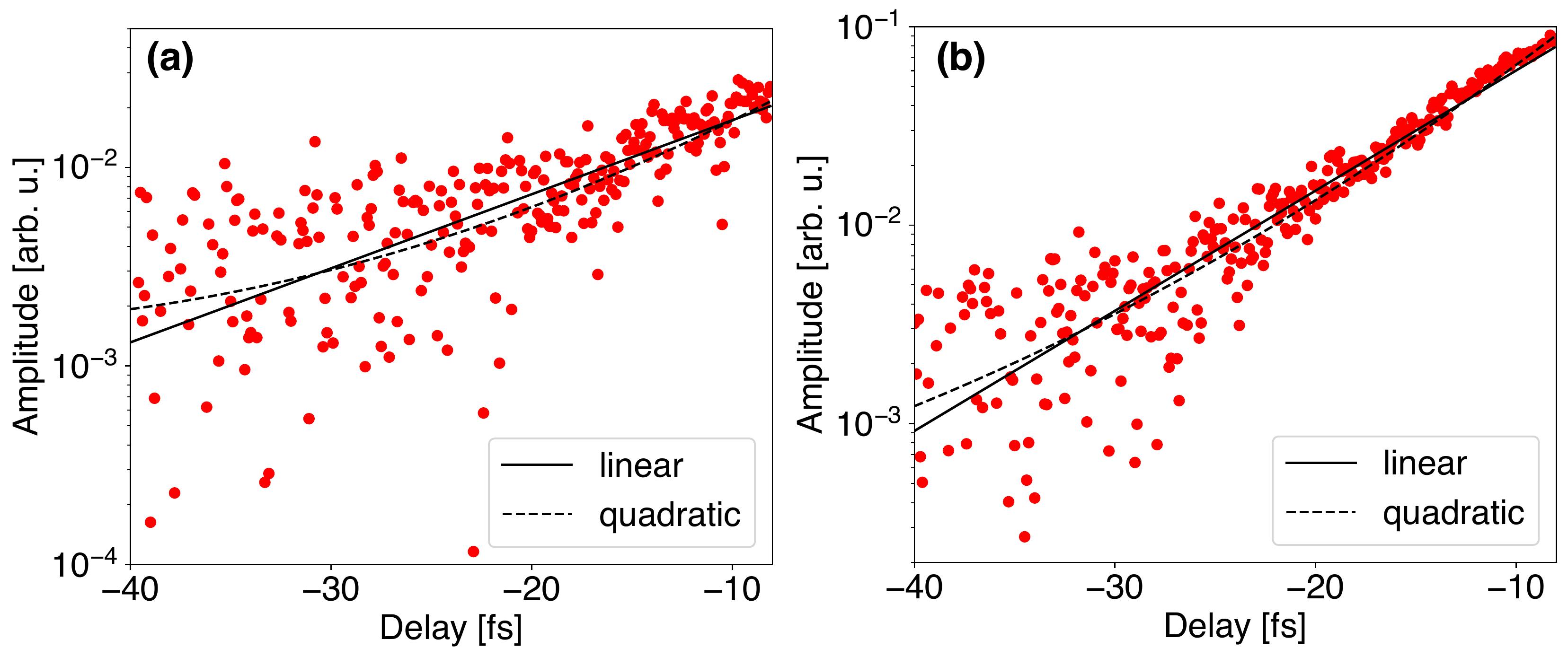}
	\caption{Quadratic fitting (dashed line) of the logarithm of the largest component in the SVD of XUV TA signal at (a) core-exciton transiton A and B and (b) core-exciton transition C. The data points from the SVD are shown as red dots. Results of linear fitting of the components are shown in black solid lines.}
	\label{fig:svd-log}
\end{figure}

To verify whether phonon-induced dephasing contributes significantly to the decay of core-excitons, we focus on the decay dynamics of the largest component in the SVD $u_1(t)$ (Fig. \ref{fig:SVD}(c) and (f), red dots) at negative delays. The largest SVD component is plotted in logarithmic scale in Fig. \ref{fig:svd-log} and the component $\log_{10} u_1(t)$ is fitted with a quadratic function $at^2+bt+c$. The fitted coefficients of the quadratic function are listed in Table \ref{tab:log}, showing that the quadratic term $a$ is two orders of magnitude smaller than the linear term $b$. In addition, the fitted $a$ are positive rather than negative as expected for a Gaussian function. This indicates that the decay of XUV TA signal at negative time delays is exponential rather than Gaussian and the effect of phonon-induced dephasing is insignificant.
\begin{table}
	\caption{Results of quadratic fitting of $\log_{10}u_1(t)$ for core-exciton A, B and core-exciton C (Fig. \ref{fig:svd-log}).}
	\label{tab:log}
	\begin{tabular}{r|cc}
		& A,B & C \\ \hline
		a & $(6\pm 2)\times 10^{-4}$ & $(5\pm 2)\times 10^{-4}$ \\ 
		b & $0.061\pm 0.008$ & $0.084\pm 0.007$ \\
		c & $-1.21\pm0.06$ & $-0.4\pm0.07$
	\end{tabular}
\end{table}

\section{Comparison with XUV total electron yield of single crystal WS$_2$} \label{app:ALS}
\begin{figure}
	\includegraphics[width=.48\textwidth]{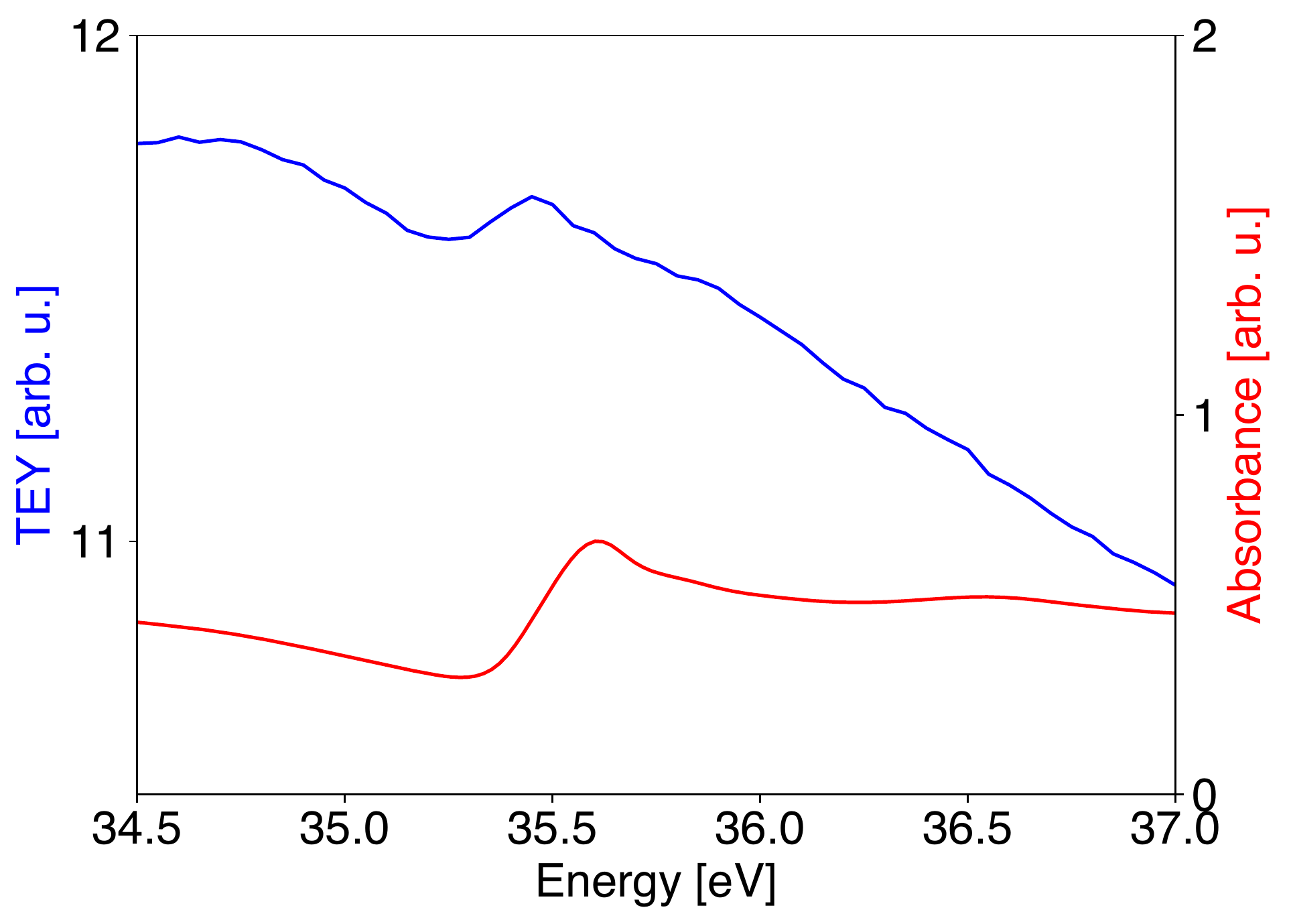}
	\caption{Normalized total electron yield spectrum of single crystal WS$_2$ measured at Beamline 4.0.3 at the Advanced Light Source (blue) and the XUV absorption spectrum of 40 nm thick WS$_2$ thin film used in the core-level transient absorption experiment (red).}
	\label{fig:BL403}
\end{figure}

To verify that the absorption peaks below 37 eV are not due to defect-induced color centers, the XUV absorption spectrum of the synthesized film is compared with total electron yield (TEY) spectrum of single crystal WS$_2$ (2Dsemiconductors USA) measured at Beamline 4.0.3 at the Advanced Light Source. The measured TEY of the single crystal sample is shown in Fig.~\ref{fig:BL403}. The spectrum is cut off at 34.5 eV due to the lack of XUV photons below 34.5 eV at the undulator beamline. The measured TEY of WS$_2$ shown in Fig.~\ref{fig:BL403} is normalized by the measured TEY of a gold film: $TEY_{norm}=TEY_{sample}/TEY_{Au}$.

\section{Additional static and transient reflectivity measurements} \label{app:reflec}
Here we provide measurements of core-exciton dynamics in the presence of photoexcited carriers in reflectivity geometry. Certain materials are challenging to synthesize as thin films for XUV absorption measurements, yet the analysis of reflectivity data alone is challenging and often relying on Kramers-Kronig transforms. Therefore, the data presented below can serve as a useful reference point for future studies of materials other than $\text{WS}_2$.

\begin{figure}
	\includegraphics[width=0.49\textwidth]{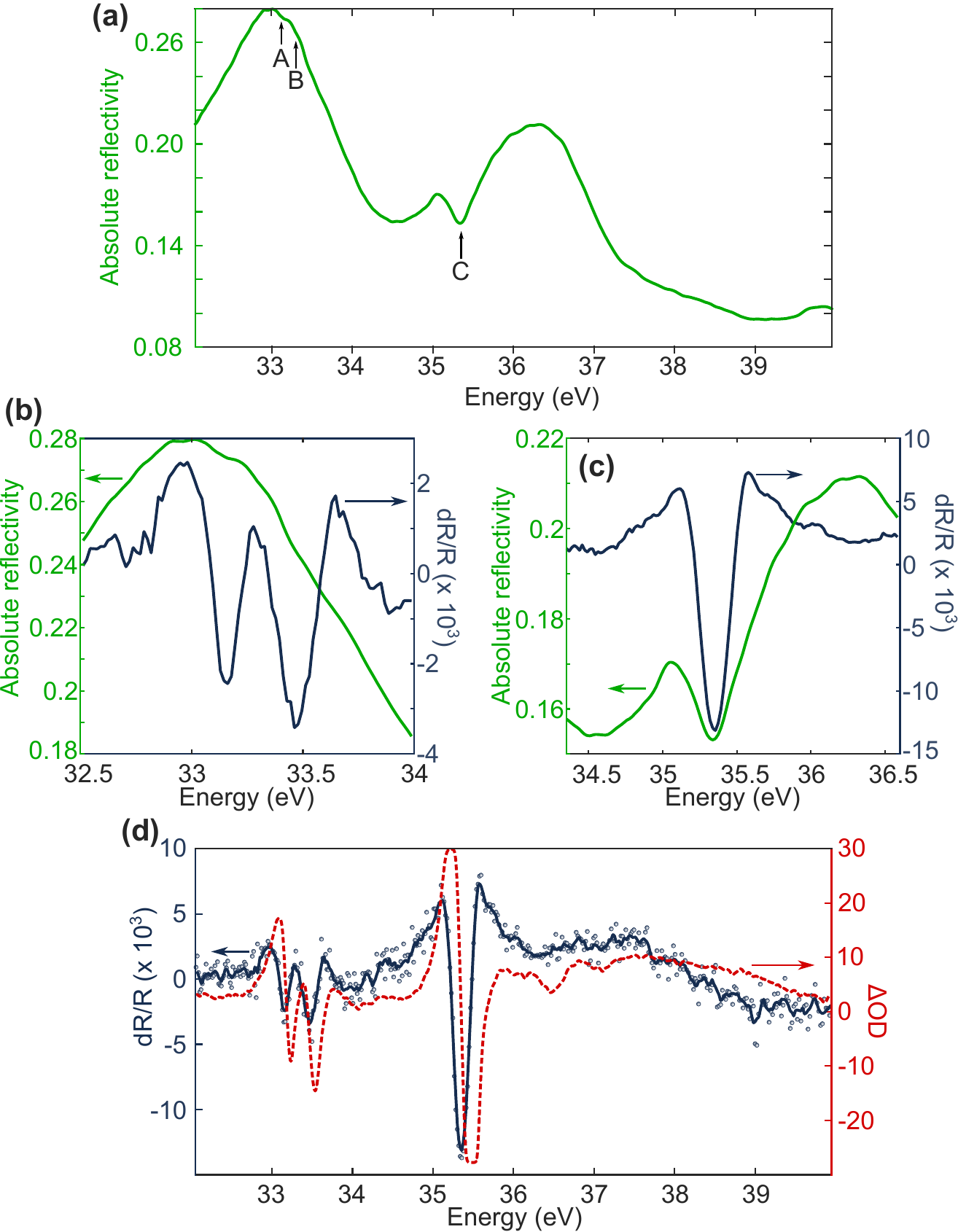}
	\caption{(a) Static XUV reflectivity ($s$-polarized) of $\text{WS}_2$. The core-exciton transitions A, B, and C are labeled. (b) and (c) present the transient reflectivity $dR/R=(R_{on}-R_{off})/R_{off}$ of core-exciton A, B, and C at $+10$ fs delay. (d) displays the changes in reflectivity at $+10$~fs (blue points), together with a 5-point moving average (blue full line), overlaid with the changes in optical density at the same pump-probe delay (red dashed line). The pump fluence used to obtain the transient reflectivity results is 25 mJ/cm$^2$. The reflectivity with and without optical pump is denoted as $R_{on}$ and $R_{off}$, respectively.}
	\label{fig:app-reflec}
\end{figure}

The absolute static reflectivity of $\text{WS}_2$ deposited on a silicon wafer, taken at 66$^\circ$ from normal (Fig. \ref{fig:app-reflec}(a)), shows that while core-exciton C is very visible, core-excitons A and B are difficult to resolve. Nevertheless, reflectivity changes are clearly observed (shown in Fig. \ref{fig:app-reflec}(b) and (c) at +10 fs delay) for each peak and share the same shape: a reduced reflectivity at the center of the exciton lineshape, and a slight increase on each side of it. The comparison (Fig. \ref{fig:app-reflec}(d)) with the transient absorption reported in the main text shows that the two observables are consistent with each other. These results display how the core-excitonic lineshapes in reflection geometry are modified by the excitation of free carriers, which has not been reported thus far.

\begin{acknowledgements}
The authors would like to thank Vincent C. Tung, Xuan Wei, Frank de Groot, Diana Y. Qiu, Felipe H. da Jornada, Han Wang
and David Prendergast for fruitful discussions. The authors would also like to thank Tevye Kuykendall for help during sample synthesis. Investigations
were supported by the Defense Advanced Research Projects Agency PULSE
Program Grant W31P4Q-13-1-0017 (concluded), the U.S. Air Force Office
of Scientific Research Nos. FA9550-19-1-0314, FA9550-20-1-0334, FA9550-15-0037
(concluded), and FA9550-14-1-0154 (concluded), the Army Research Office
No. WN911NF-14-1-0383, and the W.M. Keck Foundation award No. 046300-002.
This research used resources of the Molecular Foundry and the Advanced Light Source,
U.S. DOE Office of Science User Facilities under contract no. DE-AC02-05CH11231.
Core-level absorption simulations is conducted at Molecular Graphics and Computation
Facility, UC Berkeley College of Chemistry, funded by National Institute
of Health (NIH S10OD023532). H.-T. C. acknowledges support from Air
Force Office of Scientific Research (AFOSR) (FA9550-15-1-0037 and
FA9550-19-1-0314) and W. M. Keck Foundation (No. 046300-002); A. G. acknowledges
support from German Research Foundation (GU 1642/1-1); J. O. is supported by W. M. Keck Foundation (No. 046300-002) and Basic Science Research Program through the National Research Foundation of Korea funded by the Ministry of Education (2019R1A6A3A03032979). D. M. N. acknowledges support
from the U.S. Air Force Office
of Scientific Research (No. FA9550-15-0037, concluded) and the Army Research Office under Grant No. W911NF-20-1-0127.

\end{acknowledgements}

\bibliographystyle{apsrev4-2}

%

\end{document}